\documentclass[letterpaper,11pt,onecolumn]{article}

\usepackage[]{epsf,epsfig, amsmath,amssymb,amsthm, latexsym, color,graphicx, xspace, url}

\usepackage{times,euscript, balance}

\usepackage{xspace}

\newtheorem{theorem}{Theorem}[section]
\newtheorem{lemma}[theorem]{Lemma}
\newtheorem{proposition}[theorem]{Proposition}

\newtheorem{corollary}[theorem]{Corollary}

\def\eps{{\varepsilon}}

\def\A{\EuScript{A}}

\def\B{\EuScript{B}}

\def\E{\EuScript{E}}
\def\F{\EuScript{F}}

\def\H{\EuScript{H}}

\def\R{\EuScript{R}}
\def\S{\EuScript{S}}
\def\T{\EuScript{T}}
\def\U{\EuScript{U}}

\def\etal{\textit{et~al.}}

\def\bd{{\partial}}

\def\reals{{\mathbb R}}

\newcommand{\Rd}{{\reals}^d}
\def\bd{{\partial}}

\def\uu{{\bf u}}

\def\xx{{\bf x}}

\DeclareMathOperator{\EE}{\mathbf{Exp}}
\DeclareMathOperator{\Prob}{\mathbf{Prob}}

\newdimen\instindent
\def\institute#1{\gdef\@institute{#1}}

 \newfont{\affaddr}{phvr at 11pt}
 \newfont{\affaddrit}{phvro at 11pt} % GM 2/4/2000

\usepackage{graphicx,floatflt,psfrag}
\begin{document}

\begin{titlepage}
\title{Small-Size Relative $(p,\eps)$-Approximations for Well-Behaved Range Spaces\thanks{
    Work on this paper has been supported by NSF under grant CCF-12-16689.
  }}

\author{Esther Ezra\thanks{%
    Courant Institute of Mathematical Sciences,
    New York University, New York, NY 10012, USA;
    \textsl{esther@courant.nyu.edu}.
  } 
}

% \date{}
\maketitle

\begin{abstract}
We present improved upper bounds for the size of relative $(p,\eps)$-approximation
for range spaces with the following property: 
For any (finite) range space projected onto (that is, restricted to) a ground set of size $n$ and
for any parameter $1 \le k \le n$, the number of ranges
of size at most $k$ is only nearly-linear in $n$ and polynomial in $k$.
Such range spaces are called ``well behaved''.
Our bound is an improvement over the bound $O\left(\frac{\log{(1/p)}}{\eps^2 p}\right)$ introduced
by Li~\etal~\cite{LLS-01} for the general case (where this bound has been shown to be tight in the worst case),
when $p \ll \eps$. 
We also show that such small size relative $(p,\eps)$-approximations can be constructed in 
expected polynomial time.

Our bound also has an interesting interpretation in the context of ``$p$-nets'':
As observed by Har-Peled and Sharir~\cite{HS-11}, $p$-nets are special cases of 
relative $(p,\eps)$-approximations.
Specifically, when $\eps$ is a constant smaller than $1$, the analysis in~\cite{HS-11, LLS-01}
implies that there are $p$-nets of size $O(\log{(1/p)}/p)$ that are \emph{also} relative approximations.
In this context our construction significantly improves this bound for well-behaved range spaces. 
Despite the progress in the theory of $p$-nets and the existence of improved bounds 
corresponding to the cases that we study, these bounds do not necessarily guarantee a bounded relative error.

Lastly, we present several geometric scenarios of well-behaved range spaces, and show the resulting
bound for each of these cases obtained as a consequence of our analysis.
In particular, when $\eps$ is a constant smaller than $1$, our bound for points and axis-parallel boxes in two and 
three dimensions, as well as points and ``fat'' triangles in the plane,
matches the optimal bound for $p$-nets introduced in~\cite{AES-10, PT-11}. 
%Each of these bounds is (almost) matched with the previously known improved bounds for $p$-nets when
%$\eps$ is a constant smaller than $1$.

\end{abstract}
\end{titlepage}

\vspace{-2ex}
\section{Introduction}
\label{sec:intro}
\vspace{-2ex}

%A \emph{range space} $(X,\R)$ is a pair consisting of an underlying
%universe $X$ of objects and a family $\R$
%of subsets of $X$ (called \emph{ranges}). 

Motivated by the problem of \emph{approximate range counting}, relative $(p,\eps)$-approximations have been
introduced by Har-Peled and Sharir~\cite{HS-11}, where they revisited the seminal work of Li~\etal~\cite{LLS-01},
and showed how to apply it in order to derive known bounds on the size of various notions of sampling, and, in particular,
relative $(p,\eps)$-approximations. We recall their definition:
A range space $(\U,\R)$ is a pair consisting of an underlying universe $\U$ of objects and a certain collection $\R$ 
of subsets (\emph{ranges}). Of particular interest are range spaces of finite VC-dimension; 
the reader is referred to~\cite{HW87} for the exact definition. Informally, it suffices to require that, for any finite subset
$X \subset \U$, the number of distinct sets $\tau \cap \R$ for $\tau \in \R$ be $O(|X|^d)$ for some 
constant $d$ (which is upper-bounded by the VC-dimension).
%
%We are given a range space $(X,\R)$, where $X$ is a set of $n$ objects and $\R$ is a collection of subsets of $X$,
%called \emph{ranges}. In a typical geometric setting $X$ is a subset of some infinite ground set $\U$,
%e.g., $\U = \Rd$, for some low or fixed dimension $d$, $X$ is a finite point set in $\Rd$, 
%and $\R = \{ \tau \cap X \mid \tau \in \R_{\U} \}$,
%where $\R_{\U}$ is a collection of subsets (ranges) of $\U$ of some simple shape, such as halfspaces, simplices, balls,
%etc. In ``dual'' range spaces, we flip the roles of $X$ and $\R$, yielding a range space
%$(\R, X^*)$, where $X^*$ is a collection of subsets of $\R$, each consisting of those ranges that contain 
%a fixed element of $X$.
%In typical geometric settings $\R$ is a (finite) collection of simply-shaped regions in some low-dimensional space 
%$\Rd$ (as above),
%and $X^*$ is a collection of subsets of $\R$, each consisting of those regions that contain (i.e., ``stabbed'' by)
%a fixed point of $\Rd$.
%
%We make the standard assumption that the range space $(X,\R)$ (or, in fact, $(\U, \R_{\U})$)
%has finite \emph{VC-dimension} $d$, which is a constant independent of $n$;
%the reader is referred to~\ref{HW87} for the exact definition. Informally, it suffices to require that, 
%for any finite subset $P \subset \U$, the number of distinct sets $\tau \cap \R_{\U}$ for $\tau \in \R$ be $O(|P|^d)$ for some 
%constant $d$ (which is upper-bounded by the VC-dimension).
This is indeed the case in many geometric applications. 
In a typical geometric setting $X$ is a subset of some infinite ground set $\U$,
e.g., $\U = \Rd$, for some low or fixed dimension $d$, and thus $X$ is a finite point set in $\Rd$, 
and %$\R = \{ \tau \cap X \mid \tau \in \R_{\U} \}$, where $\R_{\U}$ 
$\R$ is a collection of subsets (ranges) of $\U$ of some simple shape, such as halfspaces, simplices, balls,
ellipsoids, and boxes in $\Rd$, where $d$ is considered a fixed. 
%etc.
In general, range spaces involving semi-algebraic ranges of
constant description complexity, i.e., semi-algebraic sets defined as a Boolean combination of
a constant number of polynomial equations and inequalities of constant maximum degree,
have finite VC-dimension (see, e.g.,~\cite{Har-Peled-11} for further details and examples). 
%Halfspaces, balls, ellipsoids, simplices, and boxes in $\Rd$, where $d$ is considered 
%a fixed constant, are examples of ranges of this kind. 
%This assumption implies that the corresponding dual range space has a finite VC-dimension as well, see, e.g.~\cite{Har-Peled-11}.
In what follows we assume to have a finite range space defined on a set $X$ of objects, and, with a slight abuse of notation,
denote its set of ranges by $\R$.

Following the notation in~\cite{HS-11}, the \emph{measure} of a range $\tau \in \R$ is the quantity\footnote{
  Technically, $\tau \subset X$, so $|X \cap \tau| = |\tau|$. However, (i) we also use 
  this notations for subsets $Z \subset X$, and (ii) typically $X$ is a finite subset of $\Rd$ and
  the ranges are best described as intersections of simply-shaped regions with $X$, as above.}
$
\overline{X}(\tau) = \frac{|\tau \cap X|}{|X|} 
$.
%The \emph{absolute measure} of $\tau$ refers to the count $|X \cap \tau|$.
Given two parameters, $0 < p <1$ and $0 < \eps < 1$, 
we say that a subset $Z \subseteq X$ is a \emph{relative $(p,\eps)$-approximation} if it satisfies, 
for each range $\tau$,
$$
\overline{X}(\tau)(1-\eps) \le \overline{Z}(\tau) \le \overline{X}(\tau)(1+\eps) , 
\quad \mbox{if $\overline{X}(\tau) \ge p$,} \quad \mbox{and}
$$
$$
\overline{X}(\tau) -\eps p \le \overline{Z}(\tau) \le \overline{X}(\tau) + \eps p , \quad \mbox{otherwise.}
$$
In fact, a slightly more general notion is the so-called $(\nu,\alpha)$-sample~\cite{Har-Peled-11, Haussler-92, LLS-01}, 
in which case the subset $Z  \subseteq X$ satisfies, for each range $\tau$,
$$
\frac{|\overline{Z}(\tau) - \overline{X}(\tau)|}{\overline{Z}(\tau) + \overline{X}(\tau) + \nu} < \alpha .
$$
As observed by Har-Peled and Sharir~\cite{HS-11}, relative $(p,\eps)$-approximations and $(\nu,\alpha)$-samples 
are equivalent with an appropriate relation between $p$, $\eps$, and $\nu$, $\alpha$ (roughly speaking, they are equivalent
up to some constant factor). Due to this observation they conclude that the analysis of Li~\etal~\cite{LLS-01} (that shows a bound
on the size of $(\nu,\alpha)$-samples) implies that,
for range spaces of finite VC-dimension $d$,
there exist relative $(p,\eps)$-approximations of size $\frac{c d\log{(1/p)}}{\eps^2p}$,
where $c > 0$ is an absolute constant. In fact, any random sample of these many elements of 
$X$ is a relative $(p,\eps)$-approximation with constant probability. More specifically,
success with probability at least $1-q$ is guaranteed if one samples
$\frac{c d\log{(1/p)} + \log{(1/q)}}{\eps^2p}$ elements of $X$. % for a sufficiently large constant $c > 0$.
These bounds are in fact an improvement over the bound $\frac{c d\log{(1/(p\eps))} + \log{(1/q)}}{\eps^2p}$
obtained by Haussler~\cite{Haussler-92} and Pollard~\cite{Pollard-86}.

It was also observed in~\cite{HS-11} that \emph{$p$-nets} and \emph{$\eps$-approximations} are  
special cases of $(\nu,\alpha)$-samples.
%A related notion to relative $(p,\eps)$-approximations is \emph{$p$-nets} for a range space $(X,\R)$ .
The first is a subset $N\subseteq X$ with the property that any range $\tau \in \R$ with $|\tau \cap X|\ge p|X|$
contains an element of $N$, %; in other words, $N$ is a hitting set for all the ``heavy'' ranges. 
and the latter is a subset $Z \subseteq X$  with the property that any range $\tau \in \R$ satisfies:
$|\overline{X}(\tau) - \overline{Z}(\tau)| \le \eps$.

In this paper we present improved bounds on the size of relative $(p,\eps)$-approximations under certain 
assumptions, and emphasize their implications to $p$-nets---see below.
%As has been observed in~\cite{HS-11}, $p$-nets are a special case of 
%relative $(p,\eps)$-approximations when $\eps$ is taken to be a constant in $(0,1)$.

\vspace{-2ex}
\paragraph{Previous results.}
It has been shown in~\cite{VC-71} that range spaces of finite VC dimension $d$ always admit 
an \emph{absolute-error} $\eps$-approximation of size $O(\log{(1/\eps)} / \eps^2)$, where the constant of proportionality
depends on $d$ (see also~\cite{Chaz-01, Chaz-04, Mat-99, PA95}). 
In fact, a random sample of that size is an $\eps$-approximation with constant probability.
As noted in~\cite{HS-11}, the analysis in~\cite{LLS-01} (see also~\cite{Tal-94}) reduces this bound to $O(1/\eps^2)$,
where, once again, a random sample of that size is an $\eps$-approximation with a constant probability.
In fact, as shown in~\cite{Chaz-01, Chaz-04, Mat-99}, the size of the $\eps$-approximation can be
further improved to be slightly higher than $O(1/\eps^{2 -2/(d+1)})$; see~\cite{HS-11} for a more comprehensive
review of these results. 

Concerning relative $(p,\eps)$-approximations, 
as noted in~\cite{HS-11}, any absolute error $\eps p$-approximation $Z$ will
approximate ranges of measure at least $p$ to within relative error $\eps$. Nevertheless,
the bound in~\cite{VC-71} for absolute-error approximations just mentioned yields a sample of size 
$O(\log{(1/(\eps p))} /\eps^2 p^2)$ in this case, whereas the bound of Li~\etal~\cite{LLS-01} is smaller by 
roughly a factor of $1/p$ (see the discussion above). 
%Specifically, Li~\etal~\cite{LLS-01} have shown a bound of 
%$O(\log{(1/\nu)}/ \alpha^2 \nu)$ on the size of $(\nu, \alpha)$-samples (see above for more details), 
%which, by the observations made in~\cite{HS-11}, implies a bound of $O(\log{(1/p)}/ \eps^2 p)$ on the size
%of relative $(p,\eps)$-approximations (in both bounds the constant of proportionality depends on $d$ and is chosen
%to be sufficiently large).

In addition to the observations made by Har-Peled and Sharir~\cite{HS-11},
their analysis also improves the bound on the size of relative $(p,\eps)$-approximations for 
several special cases of geometric range spaces. Specifically, the bound obtained for point sets in the plane 
and halfplane ranges has been improved to $O\left( \frac{\log^{4/3}{(1/(p\eps))}}{\eps^{4/3}p}\right)$ 
(this is an improvement if $\eps$ is sufficiently small with respect to $p$). 
In 3-space they reduced the bound to $O\left(\frac{\log^{3/2}{(1/(p\eps))}}{\eps^{3/2}p}\right)$, 
although this latter case is somewhat restricted, as this is the bound on the overall size of 
$O(\log{(1/p)})$ subsets, and each halfspace range has one such subset that constitutes its relative approximation.

\vspace{-2ex}
\paragraph{Our results.}
In this paper we derive improved bounds 
%(where we improve the dependency on $p$, which is different than the form derived in~\cite{HS-11}) 
for range spaces with certain favorable properties, which we call ``well-behaved''.
Our goal is to improve the dependency on $p$, which is somewhat different than the improvement obtained in~\cite{HS-11}.
That is, for any induced (finite) ranges space, the number of ranges of size $k$
is only nearly-linear in the size of the space and polynomial in $k$, where $k > 0$ is an integer parameter.
We first present our technique on abstract range spaces that satisfy these properties, 
in which case we show there exists a sample of size roughly $O((\log\log{(1/p)} + \log{(1/\eps)})/\eps^2 p)$
(see Corollary~\ref{cor:small_relative_approx} for the exact bound),
from which the relative approximation is derived.
We also present an algorithm that constructs such a sample in expected polynomial running time.
%dual range spaces of (pseudo-)disks (as objects) and points (as ranges) in the plane,
%and then show how to derive a relative approximation based on our sampling scheme.
In fact, this sample consists of two subsets of the aforementioned overall size, on which we assign weights,
in order to obtain a single \emph{weighted} sample with the desired properties;
%from which the approximation of the measure $\overline{X}(\tau)$ is derived 
see Section~\ref{sec:construction} for this derivation and the discussion below.

%\esther{I will later present a table for dual range spaces involving fat wedges, 
%  fat triangles, locally fat objects, and also primal range spaces of points and halfspaces in two
%  and three dimensions, points and disks in two dimensions, and points and axis-parallel rectangles
%  in two and three dimensions.}

As observed by Har-Peled~\cite{Har-Peled-11} relative approximations are interesting in the case where 
$p = o(\eps)$, since one can approximate ranges of measure larger than $p$ with a sample that has 
only linear dependency on $1/p$.
%their size has only a (nearly) linear dependency on $1/p$.
Our bounds improve over the bound of Li~\etal~\cite{LLS-01} for these cases.
%$1/\eps = o((1/p)^C)$,
%for any constant $C > 0$ (e.g., $1/\eps$ is at most poly-logarithmic in $1/p$). 
Nevertheless, our bounds also have an interesting interpretation when $\eps$ is some fixed constant
in $(0,1)$.
%That is, assume that $\eps$ is a constant in the range $(0,1)$. 
In this case, %as observed by Har-Peled and Sharir~\cite{HS-11},
as mentioned earlier, the relative $(p,\eps)$-approximation becomes a \emph{$p$-net}.
%This is an appropriate assumption, as for larger values of $\eps < 1$ we can use the fact that 
%in these cases 
Thus when $\eps$ is a constant smaller than $1$, a $p$-net of size $O(\log{(1/p)}/p)$ (with a constant of proportionality 
depending on the VC dimension $d$) always exists by applying the bound of Li~\etal~\cite{LLS-01}.
In this case, each range of measure at least $p$ 
%(where $n$ is the number of input objects) 
contains roughly $\log{(1/p)}$ points of the sample.
Nevertheless, in some cases
%in cases where smaller $p$-nets exist, 
it might be wasteful to sample this number of points
in order to guarantee a small relative error.
%such a property is not necessarily guaranteed if we take $p$-net of smaller size (in case 
%such a $p$-net exists).
There are several known constructions for small-size $p$-nets,
%so improved bounds for relative $(p,\eps)$-approximations are immediately obtained from 
%known constructions for small-size $p$-nets (for our case of disks and points in the plane, but also
%for more general cases), 
see, e.g.,~\cite{AES-10, CV07, Mat-92, Lauen-08, MSW90, PR08, Var-09} for several such improved bounds.
However, we are not aware of any such small-size $p$-nets that are \emph{also} relative approximations.
Our bound guarantees such properties. For example, for primal range spaces of points and axis-parallel 
boxes in two and three dimensions this bound becomes $O\left(\frac{ (\log\log{(1/p)}} {p}\right)$ instead 
of the standard bound $O(\log{(1/p)}/p)$. This bound matches the previously known (optimal) bound for $p$-nets~\cite{AES-10, PT-11}.
%(without the guarantee of satisfying the relative approximation property).

We also note that the task of constructing a relative $(p, \eps)$-approximation is much more difficult than just 
constructing $p$-nets. Indeed, in $p$-nets we aim to add sufficiently many objects (e.g., points) into the output
sample so as to guarantee that each range of measure at least $p$ is indeed stabbed, whereas in 
relative $(p, \eps)$-approximations 
%(or $p$-nets that also guarantee a bounded relative error), 
we also need to keep the set of the chosen objects under control in order to have a balance between the original measure and the 
approximate measure for each range $\tau$.

In our analysis we initially replace the set of the input objects with a relative $(p,\eps)$-approximation
$\F$ of size $O\left(\frac{\log{(1/(p\eps))}}{\eps^2p}\right)$.
We then classify the objects of $\F$ as ``heavy'' (that is, objects that participate in many ranges) and ``light''
(otherwise). We show that the number of heavy objects is relatively small.
For the light objects, 
% and then reduce this size using a sampling scheme, in which 
we randomly and independently choose each of them into the new relative 
$(p,\eps)$-approximation with some probability $\pi$ (calibrated to produce a sample of a certain expected size).
Since the original range space is well-behaved, we are able to show that each range $\tau$ (of any measure $0 < p' < 1$) 
admits a small degree of dependency in the space of events ``$\A_{\tau}$: does the relative error for $\tau$ exceeds $\eps$?''. 
We then apply the asymmetric version of the 
%We bound the probability that the relative error for a fixed range $\tau$ exceeds $\eps$ (using
%the standard Chernoff's bound), and then show that in the appropriate projected range space, the overall majority
%of objects participate in ranges that admit a small ``degree of dependence'' (see below for further details concerning this property).
%This enables us to use 
Local Lemma of Lov{\'a}sz in order to conclude that with a positive probability 
there exists such a ``good'' sample for which the relative error of all such ranges does not exceed $\eps$.
In fact, we show that both conditions of having a good sample (in the above sense) and keeping its size close 
to its expectation (up to some constant factor) happen with a positive probability---this latter condition 
can be included into the Local Lemma, which extends to that case.   
%The remaining objects (which may also involve ranges of larger degree of dependence) are handled separately, 
%in fact, we show that their overall number is relatively small, and we thus include them into the final 
%relative $(p,\eps)$-approximation. 
See Section~\ref{sec:construction} for these details.

%\esther{ Add comment A of Micha.}
The manner in which we construct our relative approximation enforces a somewhat different form 
for the measure $Z(\tau)$. This is due to two main ingredients of our analysis: 
(i) Contrary to the standard constructions of relative $(p,\eps)$-approximations~\cite{LLS-01} (as well the 
the construction in~\cite{HS-11}), absolute $\eps$-approximations (e.g.,~\cite{Chaz-01}), and $p$-nets (e.g.,~\cite{CV07, Var-09}),
where the size of the sample is fixed\footnote{In the sampling model in~\cite{LLS-01} the size of the sample
  is always bounded by $O\left(\frac{\log{(1/p)}}{\eps^2 p}\right)$ but the objects are chosen with repetitions.},
in our probabilistic model each object is chosen independently with a fixed probability $\pi$, and thus in the 
denominator of the measure we replace the size of the sample by its expectation.
(ii) Our sample consists of two subsamples, where the first one $H$ consists of a pre-determined subset of the input,
which, due to its small size, is taken in its entirety into the output, and the second sample $\F_1$ 
is obtained by choosing each input object randomly and independently with probability $\pi$ . 
This results in a biased sample.
%according to the construction of $H$ and $\F_1$. 
Specifically, we obtain a \emph{weighted} measure, in which we assign a unit weight to each object in $\F_1$,
and a weight $\pi$ to each object in $H$.

Lastly, we list several useful applications of well-behaved geometric range spaces, 
including primal range spaces of points and halfspace ranges in two and three dimensions,
points and axis-parallel boxes in two and three dimensions
(as well as points and ``fat'' triangles in the plane), and dual range spaces involving planar 
regions of nearly-linear union complexity. 
%For the latter case, our bound matches the optimal bound for $p$-nets when taking $\eps$ to be a constant smaller than 
%one~\cite{AES-10, PT-11}. 
%See Section~\ref{sec:applications} for details.

We note that our technique is inspired by the machinery of Varadarajan~\cite{Var-09} for constructing 
small-size $p$-nets for dual range spaces of ``$\alpha$-fat'' triangles and points in the plane, where
the idea for exploiting the (simpler version of the) Local Lemma of Lov{\'a}sz has initially been introduced,
as well as classifying each object as ``heavy'' or ``light''.
Nevertheless, the technique in~\cite{Var-09} does not necessarily produce a relative approximation, but only
guarantees that the sample is a $p$-net,
%each range that contains at least a $p$-fraction of the input objects has a non-empty intersection with the produced sample, 
which is the reason we had to generalize and enhance the ideas in~\cite{Var-09} 
in order to be matched with the more intricate scenario arising in our problem.

\vspace{-2ex}
\section{The Construction for Well-Behaved Range Spaces}
\label{sec:construction}

\subsection{Preliminaries}
\label{sec:prelim}

\vspace{-2ex}
\paragraph{Well-behaved range-spaces.}
Let $(\U, \R_{\U})$ be a range space of finite VC dimension. 
We say it is \emph{well behaved} it it has the following property: 
Let $(X, \R)$ be any range space projected onto a finite subset $X \subseteq \U$, 
where $\R = \{ \{\tau \cap X\} \mid \tau \in \R_{\U} \}$), and put $n := |X|$. 
Then, for any parameter $1 \le k \le n$, the number of ranges in $\R$ of size $\le k$ is at most $O(n \phi(n) k^c)$,
where $\phi(\cdot)$ is a slowly-growing function, 
%which we assume grows slower than any poly-logarithmic function, 
and $c > 0$ is an absolute constant.
In other words, for any induced (finite) range space $(X,\R)$ the number of ranges of size at most $k$ is 
only nearly-linear in $|X|$ and polynomial in $k$.
Note that by definition any induced finite range space $(X,\R)$ as above is also well-behaved\footnote{We note 
  that just the fact that $(\U, \R_{\U})$ is well-behaved already
  implies that it has a finite VC dimension.
  Indeed, by definition we have that the total number of ranges in any projected range space $(X,\R)$ 
  is polynomial in $|X|$. This implies that the so-called \emph{shattering dimension} is finite, and thus
  the VC dimension is also finite. See~\cite{Har-Peled-11} for these straightforward details.}. 

%From now on we assume we are given a finite collection $\D$ of (pseudo-)disks in the plane, and each range $\tau$ in the
%corresponding dual range space corresponds to a point (also denoted as $\tau$) in the \emph{arrangement} $\A(\D)$ 
%of $\D$ (i.e., the decomposition of the plane into vertices, edges, and faces (also referred to as \emph{cells}), 
%each being a maximal connected set contained in the intersection of a fixed subcollection of the pseudo-disks 
%of $\D$ and not meeting any other pseudo-disk), where $\D \cap \tau$ is the set of disks containing $\tau$.
%Let $P$ denote this collection of point-ranges.
%We observe right away that $\D \cap \tau$ is the same for all points $\tau$ in the same cell of $\A(\D)$,
%so with a further abuse of notation, we denote by $P$ the set of all cells of $\A(\D)$, and by $\tau$ a cell in $P$.

%\esther{The following assumption might be replaced by the assumption 

In what follows we assume, without loss of generality, that $0 < p \le 1/8$. Otherwise,
if we also have $\eps \ge 1/8$, then the size of the relative approximation is a constant, and 
if $0 < \eps < 1/8$ is arbitrary (and $p > 1/8$) then an (absolute) $(\eps/8)$-approximation always 
yields an error smaller than $\eps p$. In this case the size of the sample is only $O(1/\eps^2)$,
as shown by Li~\etal~\cite{LLS-01}.
In addition, we assume $p \le \eps$, otherwise, we output a sample of size $O(\log{(1/p)}/ \eps^2 p)$,
as shown in~\cite{LLS-01}.

%\esther{Replace it by the assumption that $\eps$ is smaller than some constant $\le 1/2$?}
  
%In what follows we assume, without loss of generality, that $\eps$ lies in the range $(0,1/4]$. 
%This is an appropriate assumption, as for larger values of $\eps < 1$ we can use the fact that 
%in these cases any \emph{$p$-net} is also a relative $(p,\eps)$-approximation (see~\cite{HS-11}
%for this rather trivial observation), so improved bounds for relative $(p,\eps)$-approximations are
%immediately obtained from 
%known constructions for small-size $p$-nets (for our case of disks and points in the plane, but also
%for more general cases), see, e.g.,~\cite{AES-10,CV07, Mat-92, MSW90 ,PR08} for several such improved bounds.

%Let $\D$ be a set of $n$ pseudo-disks, and let $P$ be the resulting collection of ranges %(or rather cell-ranges)
%induced by $\A(\D)$.

\vspace{-2ex}
\subsection{The Construction}
\label{sec:actual_construction}
\vspace{-2ex}
Let $(X,\R)$ be a well-behaved (finite) range space.
We first replace the objects in $X$ by a sample $\F \subseteq X$ that is a relative $(p,\eps)$-approximation
for $(X, \R)$; let $\T$ be the resulting collection of ranges projected onto $\F$. 
By the discussion in Section~\ref{sec:intro}, there exists such a sample of size 
$\frac{D \log{(1/(p \eps))}}{\eps^2p}$, where $D > 0$ is a sufficiently large absolute constant.
Moreover, a random sample of $X$ of that size is a relative $(p,\eps)$-approximation with constant probability.\footnote{
  Note that this bound is somewhat suboptimal, as the $\log{(1/\eps)}$ factor in the enumerator can be removed by the analysis
  of~\cite{LLS-01}. However, due to technical reasons imposed by our analysis, we need a slightly larger sample, which is the size
  that we choose.
}
%In fact, for technical reasons imposed by our analysis, we need a slightly larger sample of size 
%$\frac{C \log{(1/(p \eps))}}{\eps^2p}$, for a sufficiently large absolute constant $C > 0$.
The replacement of $X$ with $\F$ implies that each range $\tau \in \R$ satisfies:
\begin{equation*}
  %\label{eq:approx1}
 \frac{|\tau \cap X|}{|X|}(1-\eps) 
 \le \frac{|\tau \cap \F|}{|\F|} 
 \le  \frac{|\tau \cap X|}{|X|}(1+\eps) \quad \mbox{if $\overline{X}(\tau) \ge p$,}
\end{equation*}
and
$$
\frac{|\tau \cap X|}{|X|} - \eps p 
\le \frac{|\tau \cap \F|}{|\F|} 
\le  \frac{|\tau \cap X|}{|X|} + \eps p \quad \mbox{otherwise.}
$$

%For the time being, we only consider point-ranges $\tau$ with 

From now on, we focus on the construction of an improved relative $(p,\eps)$-approximation for $(\F,\T)$.
This (standard) reduction involves no loss of generality, because, as is easily verified, a relative 
$(p(1 - \eps),\eps)$-approximation
of a relative $(p,\eps)$-approximation is a relative $(p,2\eps + \eps^2)$-approximation.
Hence by scaling $p$ and $\eps$ by the appropriate constant factors, it suffices to construct a relative
$(p,\eps)$-approximation for $(\F,\T)$.
%\esther{I scaled $\eps$ as it can take any value between $0$ and $1$. }

Our construction partitions $\tau \cap \F$  into two subsets, and represents
the approximation for $|\tau \cap \F|$ as the sum of two appropriate sample measures---see below
for details, and for the consolidation of the two samples into a common (weighted) set. 

A range $\tau$ is said to lie at the \emph{$i$th layer} $\S_i$ of $\T$, for $i = 1, \ldots, \log{(1/p)}$,
if it satisfies 
\begin{equation}
  \label{eq:layer}
  2^{i-1} p \le \frac{|\tau \cap \F|}{|\F|} < 2^{i}p .
\end{equation}
For the sake of completeness, the $0$th layer $\S_0$ consists of those ranges $\tau$ with 
$\frac{|\tau \cap \F|}{|\F|} < p$.

Consider a fixed layer $\S_i$, where $i \ge 1$.
Put $\Delta_i := \frac{C 2^{i-1} \log{(1/(p\eps))}}{\eps^2}$.
Equation~(\ref{eq:layer}) and the bound on $|F|$ then imply that each range $\tau \in \S_i$ satisfies
$\Delta_i  \le |\tau \cap \F|  \le 2 \Delta_i $.
%
%\Delta_i (1-\eps) \le |\F| \cdot \frac{|\tau \cap X|}{|X|} (1-\eps)  \le 
%|\tau \cap \F| \le |\F| \cdot \frac{|\tau \cap X|}{|X|} (1+\eps) \le 2 \Delta_i(1+\eps) .
If $\tau \in \S_0$, then Equation~(\ref{eq:layer}) implies
$
0 \le |\tau \cap \F| < \Delta_1 = 2\Delta_0
%0 \le |\tau \cap \F| \le |\F| \cdot \left( \frac{|\tau \cap X|}{|X|} + \eps p\right) \le 
%|\F| p (1+\eps) = \Delta_1 (1+\eps) = 2\Delta_0 (1+\eps).
$.

In other words, all ranges $\tau \in \S_i$ have size at most $2 \Delta_i$ 
(and at least $\Delta_i$, if $i \ge 1$),
%in the arrangement $\A(\F)$ of $\F$, 
$i=0, \ldots, \log{(1/p)}$, and they appear now as ranges from the range space $(X,\R)$ projected onto (that is, restricted to) $\F$.
The assumption that the (original) range space $(X,\R)$ is well-behaved implies that the number of the
distinct ranges in a fixed layer $\S_i$ is only $O(|\F| \phi(|\F|) {\Delta_i}^c)$.
%The random sampling theory of Clarkson and Shor~\cite{CS-89} and the fact that the complexity of the 
%union of any set of (pseudo-)disks is linear~\cite{KLPS-86} imply that the number of such distinct cell-ranges
%is only $O(|\F| \Delta_i)$ .
%% (recall that $0 < \eps \le 1/4$).

\vspace{-2ex}
\paragraph{A classification of the objects.}
The discussion above implies that the number of containments between the objects in $\F$ and the ranges in layer $\S_i$ 
%of size at most $2 \Delta_i$ 
is $O(|\F|\phi(|\F|) {\Delta_i}^{c+1})$, for $i=0,\ldots, \log{(1/p)}$. 
We say that an object in $\F$ is \emph{heavy} in $\S_i$ if it appears in at least $A \cdot \phi(|\F|) {\Delta_i}^{c+2}$ ranges 
(at layer $\S_i$), for a sufficiently large constant $A > 0$ that depends on the constant of proportionality in the bound
on the number of containments.
Otherwise, this object is said to be \emph{light} in $\S_i$.  
%We claim 
An easy variant of Markov's inequality implies that the number of heavy objects (in $\S_i$) is at most 
$O(|\F| / {\Delta_i}) = O\left(\frac{1}{2^{i-1} p}\right)$. 
%which, in particular, is upper bounded by 
%$O\left(\frac{1/p}{\log{(1/p)}} \right)$. 
%Indeed, let $H \subseteq \F$ denote the subset of heavy disks.
%Our claim follows from the fact that the number of incidences between $H$ and the cells is at least $|H| \Delta^4$, and it is 
%always bounded from above by $O(|\F| \Delta^2)$ (as observed above).
%
Hence, the overall number of heavy objects, over all layers $i=0, \ldots, \log{(1/p)}$,
is only $O(1/p)$.  Let $H$ denote this subset, which, from now on, 
we just refer to as the \emph{heavy objects}.
%We thus put all of them in the target relative $(p,\eps)$-approximation $A$,
%and thus $|\tau \cap H| = |\tau \cap A|$.
%
Note that, by construction, each remaining object (an object in $\F \setminus H$) is light in each layer 
$i=0, \ldots, \log{(1/p)}$. Put $L := \F \setminus H$ and refer to its elements as the \emph{light objects}. 
%Denote by $H$ (resp., $L$) the set of heavy (resp., light) disks in $\F$.

We next consider, for each range $\tau$, the two subsets $H \cap \tau$, $L \cap \tau$ of heavy and light objects, 
respectively, which $\tau$ stabs, and approximate each of their measures in turn.
%with respect to the corresponding sets $H$, $L$.
%Specifically, let $L$ be the set of the light disks in $\F$.
%By definition, $L = \F \setminus H$, $L \cap H = \emptyset$, 
%and thus $|\tau \cap \F| = |\tau \cap L| + |\tau \cap H|$.
%We approximate each of these two summands in turn.
First, since the number of heavy objects is only $O(1/p)$, we put all of them 
in the target relative $(p,\eps)$-approximation.
% $A$ and thus $|\tau \cap H| = |\tau \cap A|$.
In the sequel we describe our approximation with respect to the light objects.

%\paragraph{Approximating $|\tau \cap L|$.}
\vspace{-2ex}
\subsubsection{The Analysis for the Light Objects}
\vspace{-2ex}
We first observe that the analysis above regarding the size of $|H|$ allows us to assume that $|L| \ge |\F|/2$.
Indeed, this easily follows by choosing the constant $A$ sufficiently large.
%by our assumption that $p \le 1/8$.
%provided that $\eps$ is smaller than some constant threshold (which certainly holds under our assuption that $\eps \le 1/4$).

We now restrict the range space $(\F, \T)$ to $L$.
% and let $\T_{L}$ be the set of ranges of $\T$ projected onto $L$.
Our goal is now to approximate $\overline{L}(\tau)$.
We keep associating these ranges with the \emph{same layers} $\S_i$, as defined in~(\ref{eq:layer}). 
Note that a range $\tau$ at layer $i$ may now satisfy $|\tau \cap L| \ll \Delta_i$.
%Here too we classify the ranges $\tau \in \T_{L}$ according to layers, that is, $\tau$ lies at the $i$th layer,
%$i=1, \ldots, \log{(1/p)}$, if we have a similar condition as in~(\ref{eq:layer}), where we replace the 
%measure with $\overline{L}(\tau)$, and at the $0$th layer we have $\overline{L}(\tau) < p$. 
%With a slight abuse of notation, we continue to denote these layers by $\S_i$, $i=0, \ldots, \log{(1/p)}$.  
%Note that an original range $\tau \in \T$ that lies at layer $i$ may now lie at a different layer $j$. 
This can happen, for example, when the overall majority of objects in 
$\tau \cap \F$ are heavy and then $|\tau \cap L|$ is considerably smaller than $|\tau \cap \F|$. 
%Also, observe that $|L| \ge |F|/2$, which implies we always have $\overline{L}(\tau) \le 2 \overline{F}(\tau)$.
These differences, however, do not affect the final approximation for $\overline{F}(\tau)$, which is a key observation 
in our analysis; see Section~\ref{sec:derivation} for this derivation.

We next sample each object in $L$ independently with probability 
$\pi := \frac{\max\{\log\log{(1/p)}, \log{\phi(|F|)} \} + \log{(1/\eps)}}{\log{(1/p)} + \log{(1/\eps)}}$, and
let $\F_1$ be the resulting sample. Thus its expected size is
$O\left(\frac{\max\{\log\log{(1/p)}, \log{\phi(|F|)} \}  + \log{(1/\eps)}}{\eps^2 p}\right)$.
Note that, by assumption $\phi(\cdot)$ is a sublinear function, and
$p \le \eps$, thus we always have $\pi < 1$, as is easily verified.
The main ingredients of the analysis are shown in the following proposition:

\begin{proposition}
  \label{prop:F_1} 
  With some positive probability, $\F_1$ satisfies, for every layer $\S_i$ and for every range $\tau$ 
  of $\S_i$, 
  $i = 1, \ldots, \log{(1/p)}$, (where $\EE(\cdot)$ denotes expectation)\footnote{Note that due to our 
    sampling model we replace $|\F_{1}|$ by $\EE\{\F_1\}$, which is in fact $\pi \cdot |L|$.}:
  \begin{align}
    \label{eq:approx_size}
    \frac{|\tau \cap L|}{|L|}\cdot(1-\eps)  \le \frac{|\tau \cap \F_{1}|}{ \EE\{\F_{1}\} } \le \frac{|\tau \cap L|}{|L|}\cdot (1+\eps) , &
    \quad \mbox{if $|\tau \cap L| \ge 2^{i-1}p |\F|$,} 
  \end{align}
  and
  \begin{align*}
    \frac{|\tau \cap L|}{|L|} - \eps \cdot (2^{i-1}p) \le \frac{|\tau \cap \F_{1}|}{ \EE\{\F_{1}\} } \le \frac{|\tau \cap L|}{|L|} + \eps \cdot (2^{i-1}p) ,
    & \quad \mbox{otherwise.}
  \end{align*}
  When $i=0$ we have for each $\tau \in \S_0$ (in which case $|\tau \cap L| < p|\F|$):
  $$
  \frac{|\tau \cap L|}{|L|} - \eps p \le \frac{|\tau \cap \F_{1}|}{ \EE\{\F_{1}\} } 
  \le \frac{|\tau \cap L|}{|L|} + \eps p .
  $$
\end{proposition}
\begin{proof}
Fix a layer $\S_i$ and a range $\tau \in \S_i$, $i \ge 1$.
%\esther{I think the same analysis we had for $i \ge 1$ should also hold for $i=0$.}
%We first consider the case $i \ge 1$, and then resort to the case $i=0$.
Let $\A_{\tau}$ be the event that $\F_1$ does not satisfy~(\ref{eq:approx_size}) for $\tau$.
%the relative $(p,\eps)$-approximation property for $\tau$. 
We consider separately the following two cases:
$|\tau \cap L| \ge 2^{i-1}p |\F|$, and $|\tau \cap L| < 2^{i-1}p|\F|$.

\noindent{\bf (i) $|\tau \cap L| \ge 2^{i-1}p|\F|$:}
Since $\EE\{|\F_1|\} = \pi |L|$, the event $\A_{\tau}$ in this case is
$$
\frac{|\tau \cap \F_1|}{\pi |L|} < \overline{L}(\tau)(1-\eps) \quad \mbox{or} \quad
\frac{|\tau \cap \F_1|}{\pi |L|} > \overline{L}(\tau)(1+\eps) .
$$
That is, 
$|\tau \cap \F_1| < \pi |\tau \cap L|(1-\eps)$ \quad or \quad $|\tau \cap \F_1| > \pi |\tau \cap L|(1+\eps)$.
%\esther{One problem is that I need to replace the measure $\overline{L_1}(\tau)$ with 
%  $\frac{|\tau \cap L_1|}{\pi |L|}$, as $L_1$ is a sample chosen with repetitions and thus $|L_1|$
%  is not a fixed amount, but a random variable whose expectation is $\pi |L|$. It will be interesting to show that 
%  in both cases the resulting probability is more or less the same.}
Since $\EE\{|\tau \cap \F_1| \} = \pi \cdot |\tau \cap L|$, %(where $\EE(\cdot)$ denotes expectation),
we have:
$$
\Prob\{\A_{\tau}\} = \Prob\bigl\{|\tau \cap \F_1| < \EE\{|\tau \cap \F_1| \}(1-\eps) \bigr\} + 
                    \Prob\bigl\{|\tau \cap \F_1| > \EE\{|\tau \cap \F_1| \}(1+\eps) \bigr\} .
$$
Using Chernoff's bound (see, e.g.,~\cite{AS-00}) and the fact that $|\tau \cap L| \ge |\F|\cdot 2^{i-1} p$, 
we thus obtain:
%\begin{align*}
{\small
$$
  \Prob\{\A_{\tau}\} < 2 \exp\left\{- \eps^2 \EE\{|\tau \cap \F_1|\} /3 \right\}  %\\
  = 2 \exp\left\{- \pi \eps^2 |\tau \cap L| /3 \right\} 
  \le 2 \exp\left\{- \pi \eps^2 \cdot 2^{i-1}p |\F| /3 \right\} .
%\end{align*}
$$
}
%Since we assume $|L| \ge |\F|/2 = \frac{C \log{(1/(p\eps))}}{2 \eps^2 p}$, 
Substituting $|\F| = \frac{D \log{(1/(p\eps))}}{\eps^2 p}$, 
$\pi = \frac{\max\{\log\log{(1/p)}, \log{\phi(|\F|)}\} + \log{(1/\eps)}}{\log{(1/p)} + \log{(1/\eps)}}$, we obtain
$$
\Prob\{\A_{\tau}\} < 2 \exp\left\{- \frac{D}{3} \cdot 2^{i-1} \cdot (\max\{\log\log{(1/p)}, \log{\phi(|\F|)}\} + \log{(1/\eps)}) \right\} < 
\left(\frac{\eps}{\log{(1/p)} \phi(|\F|)}\right)^{B 2^{i-1}} ,
$$
where $B > 0$ is a constant that depends linearly on $D$, and can be made arbitrarily large by 
choosing $D$ sufficiently large. 
%The latter inequality follows by the choice of $\pi$.

\noindent{\bf (ii) $|\tau \cap L| < 2^{i-1}p |\F|$:}
Here $\A_{\tau}$ is the event:
$$
\frac{|\tau \cap \F_1|}{\pi |L|} < \overline{L}(\tau) - 2^{i-1} \eps p \quad \mbox{or} \quad
\frac{|\tau \cap \F_1|}{\pi |L|} > \overline{L}(\tau) + 2^{i-1} \eps p.
$$
That is, 
$|\tau \cap \F_1| < \pi |\tau \cap L| - 2^{i-1}\eps p \pi |L|$ \quad or \quad 
$|\tau \cap \F_1| > \pi |\tau \cap L| + 2^{i-1} \eps p \pi |L|$.
We then have:
%\begin{align*}
{\small 
  $$
\Prob\{\A_{\tau}\} = 
\Prob\bigl\{|\tau \cap \F_1| < \EE\{|\tau \cap \F_1| \} - 2^{i-1}\eps p \pi |L| \bigr\}  
+ \Prob\bigl\{|\tau \cap \F_1| > \EE\{|\tau \cap \F_1| \} + 2^{i-1}\eps p \pi |L| \bigr\} 
$$
$$  
= \Prob\left\{|\tau \cap \F_1| < \EE\{|\tau \cap \F_1| \} \left(1 - \frac{2^{i-1} \eps p |L|}{ |\tau \cap L|}\right) \right\}   
+ \Prob\left\{|\tau \cap \F_1| > \EE\{|\tau \cap \F_1| \} \left(1 + \frac{2^{i-1} \eps p |L|}{ |\tau \cap L|}\right) \right\} .
$$
}
%We then have:
%\begin{align*}
%\Prob\{\A_{\tau}\}  
%  &
%  = \Prob\bigl\{|\tau \cap \F_1| < \EE\{|\tau \cap \F_1| \} - 2^{i-1}\eps p \pi |L| \bigr\} \\
%  & 
%  + \Prob\bigl\{|\tau \cap \F_1| > \EE\{|\tau \cap \F_1| \} + 2^{i-1}\eps p \pi |L| \bigr\} \\
%  &  
%  = \Prob\left\{|\tau \cap \F_1| < \EE\{|\tau \cap \F_1| \} \left(1 - \frac{2^{i-1} \eps p |L|}{ |\tau \cap L|}\right) \right\} \\
%  &  
%  + \Prob\left\{|\tau \cap \F_1| > \EE\{|\tau \cap \F_1| \} \left(1 + \frac{2^{i-1} \eps p |L|}{ |\tau \cap L|}\right) \right\} .
%\end{align*}
Applying once again Chernoff's bound, we obtain:
$$
\Prob\{\A_{\tau}\} 
< 2 \exp\left\{- \left( \frac{2^{i-1} \eps p |L|}{|\tau \cap L|} \right)^2 \cdot \pi |\tau \cap L|/3 \right\}
= 
2 \exp\left\{- \frac{\pi \cdot \eps^2 \cdot (2^{i-1} p)^2 \cdot |L|^2 }{3 |\tau \cap L|} \right\} .
%\le 2 \exp{- \eps^2/2 \cdot (\pi \cdot |L| p/2) \right\}
$$
Since in this case $|L \cap \tau| < 2^{i-1} p \cdot |\F|$, and, as stated above, $|L| \ge |\F|/2$,
we obtain that the latter term is bounded by
$
2 \exp\left\{- \pi \cdot \eps^2 \cdot (2^{i-1} p) \cdot |\F|/12 \right\} 
$,
on which we can derive  
%is exactly 
the same bound as in case (i), using similar considerations. Hence in summary we obtain in both cases 
%and since $|L| \ge |\F|/2$, we obtain:
%$$
%\Prob\{\A_{\tau}\} < 2 \exp\left\{- C/12 \cdot 2^{i-1} \cdot (\log\log{(1/p)} + \log{(1/\eps)})\right\} < \left(\frac{\eps}{\log{(1/p)}}\right)^{B 2^{i-1}} ,
%$$
%where $B$ is the same constant as in case (i).
%
%To summarize, in both cases we obtain:
\begin{equation}
  \label{eq:chernoff}
  \Prob\{\A_{\tau}\} < \left(\frac{\eps}{\log{(1/p)} \phi(|\F|)}\right)^{B 2^{i-1}}  .
\end{equation}

The case $i=0$ (or rather $\tau \in \S_0$) follows by similar considerations as in case (ii) above when replacing
$2^{i-1} p$ with $p$.
This yields the bound $\Prob\{\A_{\tau}\} < \left(\frac{\eps}{\log{(1/p)} \phi(|F|) }\right)^{B}$ in this case.

%When $i = 0$, we have $\overline{\F}(\tau) < p$. Using the assumption that $|L| \ge |\F|/2$,
%we obtain 
%$$
%\overline{L}(\tau) = \frac{|\tau \cap L|}{|L|} \le \frac{|\tau \cap F|}{|\F|/2} < 2p . 
%$$  

\vspace{-2ex}
\paragraph{Applying the Asymmetric Local Lemma of Lov{\'a}sz.}

We next apply the Local Lemma of Lov{\'a}sz (see, e.g.,~\cite{AS-00}), to show that 
$$
\bigwedge_{i=0}^{\log{(1/p)}} \bigwedge_{\tau \in \S_i } (1 - \Prob\{\A_{\tau}\}) > 0 .
$$
This will imply that there exists a sample $\F_1$ of $L$ that approximates $|\tau \cap L|$ as in~(\ref{eq:approx_size}), 
for all ranges $\tau$.
Specifically, we are going to apply the asymmetric version of the Local Lemma, stated below in the context of our problem.

We first observe that for a pair of ranges $\tau$, $\tau'$, the corresponding events $\A_{\tau}$, 
$\A_{\tau'}$ are mutually independent if and only if there is no object in $L$ that participates in both 
$\tau$, $\tau'$. Indeed, since we sample each object of $L$ independently, %to be included into the sample,
the two corresponding events $\A_{\tau}$, $\A_{\tau'}$ can affect each other only if there is an object in $L$ 
that $\tau$, $\tau'$ share.
In what follows we denote a pair $E$, $E'$ of mutually dependent events by $E \sim E'$.

Let $\E_i$ denote the collection of events $\A_{\tau}$ for all ranges $\tau$ at a fixed layer $i$, and
let $\E$ denote the entire collection $\bigcup_{i=0}^{\log{(1/p)}} \E_i$.
In order to apply the asymmetric version of the Local Lemma we need to show there exists an assignment 
$\xx : \E \rightarrow (0,1)$, 
such that
\begin{equation}
  \label{eq:assignment}
  \Prob\{\A_{\tau}\} \le \xx(\A_{\tau}) \cdot \prod_{\A_{\tau'} \sim \A_{\tau}, \A_{\tau'} \neq \A_{\tau}} (1 - \xx(A_{\tau'})) , 
\end{equation}
for each $\A_{\tau} \in \E$. 
The Local Lemma of Lov{\'a}sz then implies that
\begin{equation}
  \label{eq:all_events}
  \bigwedge_{A_{\tau} \in \E} (1 - \Prob\{\A_{\tau}\}) \ge \prod_{A_{\tau} \in \E} (1 - \xx(\A_{\tau})) > 0 .
\end{equation}
(Once again, see~\cite{AS-00} for further details.)
In Lemma~\ref{lem:local_lemma} we show that there exists such a valid assignment. 
This will complete the proof of the proposition. 
\end{proof}

%We next show:
\begin{lemma}
  \label{lem:local_lemma}
  The assignment $\xx(\A_{\tau}) = \exp\left\{2^{i+1}\right\} \cdot \Prob\{\A_{\tau}\}$, 
  for each $\tau \in \S_i$ and for each layer $\S_i$, $i=0, \ldots, \log{(1/p)}$,
  satisfies~(\ref{eq:assignment}) for every $\A_{\tau} \in \E$.
\end{lemma}

We postpone the proof of Lemma~\ref{lem:local_lemma} to Appendix~\ref{app:proof_local_lemma}, but leave several remarks below.

\noindent{\bf Remarks:}
\noindent {\bf 1).}
We note that for each $\tau \in \S_j$, the exponent $2^{j-1}$ in the bound on $\Prob\{\A_{\tau}\}$ ``beats'' the term $\Delta_j^{c+2}$ 
in the degree of dependency (see Appendix~\ref{app:proof_local_lemma}). 
Nevertheless, when $i > j$, this exponent cannot beat $\Delta_i$, which is the reason we
set $\xx(A_{\tau}) = \exp{2^{i+1}}\Prob\{\A_{\tau}\}$. This also demonstrates the crucial property 
%In the analysis of Proposition~\ref{prop:F_1} it is crucial to 
of classifying the ranges according to their layers, and then bounding the probability to fail to produce a relative 
$(p,\eps)$-approximation in each of these layers, as we did in Proposition~\ref{prop:F_1}.
In other words, just the information $|\tau \cap L| \ge p |\F|$ or $|\tau \cap L| < p |\F|$ (which is the standard ``cut-off'' 
in relative approximations) is insufficient to produce a relative bounded error in the manner that we do.
%
%Even though a range $\tau$, with $|\tau \cap L| < 2^{i-1}p |\F|$, may satisfy $|\tau \cap L| \ge p|\F|$, 
%and then the first part of~(\ref{eq:approx_size}) will be satisfied for $\tau$ with some probability, this may fail when applying
%the local lemma to all ranges $\tau$ (over all layers). In this case the fact that $\tau$ lies at layer $i$ implies that 
%the degree of dependency may be too large to be ``beaten'' by $\Prob\{\A_{\tau}\}$. This is the reason we always need to keep
%the exponent in $\Prob\{\A_{\tau}\}$ to be dependent on $i$.
\newline\noindent {\bf 2).}
We note that applying the simpler version of the Local Lemma of Lov{\'a}sz, in each \emph{fixed layer $i$}, 
is almost immediate (a similar step has been taken in~\cite{Var-09} for ``fat'' triangles in the plane and points). 
Indeed, 
%since each point-range $\tau$ is contained in $O(\Delta_i)$ light disks, and each such disk
%contains at most $O({\Delta_i}^3)$ cells (point-ranges), it follows that 
each range has a degree-of-dependency $\delta$ that is at most $O(\phi(|\F|) {\Delta_i}^{c+3})$. 
Following Inequality~(\ref{eq:chernoff}), we obtain $\delta \cdot \Prob\{\A_{\tau}\} < 1/e$, 
for a sufficiently large choice of $D$ (and thus of $B$). 
In this case, the Local Lemma implies that, with a positive probability, all the complementary events 
$\overline{\A_{\tau}}$ (for $\A_{\tau}$ in layer $i$) are satisfied. 
Nevertheless, this property is not guaranteed for the entire set of events over \emph{all} layers, as the interaction
among events from different layers may involve a higher degree of dependency. This is the main reason we had to resort 
to the asymmetric version of the Local Lemma, which, as our analysis shows, overcomes this difficulty,
and eventually obtains a \emph{single} sample for all layers.
%This implies that there is a sample $\F_1$, which well approximates $|\tau \cap L|$ (in the above 
%sense), for each point-range $\tau$.

\vspace{-2ex}
\paragraph{Bounding the size of the sample.}
As noted above, the expected size of $\F_1$ is $O\left(\frac{\max\{\log\log{(1/p)}, \log{\phi(|\F|} \} +  \log{(1/\eps)}}{\eps^2 p}\right)$.
Nevertheless, we need to show that a sample of that actual size exists 
%Our next goal is to show that there exists a sample $\F_1$ 
and that it satisfies the assertions in Proposition~\ref{prop:F_1}.
% and (ii) its (actual) size is $O\left(\frac{\log\log{(1/p)} +  \log{(1/\eps)}}{\eps^2 p}\right)$.
In Appendix~\ref{app:proof_sample_size} we show that Lemma~\ref{lem:local_lemma} can be extended to include
the event $\B$ that $|\F_1|$ deviates from its expectation by some constant factor. This yields:

\begin{corollary}
  \label{cor:F_1}
  There exists a sample $\F_1 \subseteq L$ that satisfies the assertion of Proposition~\ref{prop:F_1},
  whose size is $O\left(\frac{\max\{\log\log{(1/p)},\log{\phi(|\F|)} \} + \log{(1/\eps)}}{\eps^2 p}\right)$.
\end{corollary}

\vspace{-2ex}
\subsection{Deriving the Relative Approximation}
\label{sec:derivation}
\vspace{-2ex}
We now combine the two samples $\F_1$ and $H$ that we have constructed in order to derive the relative approximation
for each range $\tau$.
%We begin with the relative approximation with respect to $\frac{|\tau \cap \F|}{|F|}$.
By construction, $|\tau \cap \F| = |\tau \cap L| + |\tau \cap H|$.
%According to Proposition~\ref{prop:F_1}, 
Combining the two cases in~(\ref{eq:approx_size}) for each range $\tau$ in $\S_i$, $i=1, \ldots, \log{(1/p)}$,
and adding the term $|\tau \cap H|/|L|$ for each side of the inequality, 
we have: 
$$
\frac{|\tau \cap L|}{|L|}\cdot(1-\eps) -\eps \cdot (2^{i-1}p) + \frac{|\tau \cap H|}{|L|} 
\le \frac{|\tau \cap \F_{1}|}{ \EE\{\F_{1}\} } + \frac{|\tau \cap H|}{|L|} 
\le \frac{|\tau \cap L|}{|L|}\cdot (1+\eps) + \eps \cdot (2^{i-1}p) + \frac{|\tau \cap H|}{|L|} .
%|\tau \cap L| \cdot(1-\eps) + |\tau \cap H|  \le 
%\frac{|\tau \cap \F_1|}{\pi} + |\tau \cap H| 
%\le |\tau \cap L|\cdot (1+\eps) + |\tau \cap H|, 
%\quad \mbox{if $\overline{L}(\tau) \ge 2^{i-1}p$,}
$$
%and
%$$
%|\tau \cap L| + |\tau \cap H| - |L|\eps \cdot (2^{i-1}p) \le 
%\frac{|\tau \cap \F_1|}{\pi } + |\tau \cap H| 
%\le |\tau \cap L| + |\tau \cap H| + |L|\eps \cdot (2^{i-1}p) ,
%\quad \mbox{otherwise.}
%$$ 
Substituting $\EE\{\F_{1}\} = \pi |L|$, we obtain:
$$
|\tau \cap \F| - |\tau \cap L|\eps - \eps \cdot (2^{i-1}p) |L|
\le \frac{|\tau \cap \F_1| + \pi|\tau \cap H|}{\pi} 
\le |\tau \cap \F| + |\tau \cap L|\eps + \eps \cdot (2^{i-1}p) |L| .
% \quad \mbox{if $\overline{L}(\tau) \ge 2^{i-1}p$,}
$$
%and
%$$
%|\tau \cap \F| - |L|\eps \cdot (2^{i-1}p) 
%\le \frac{|\tau \cap \F_1| + \pi |\tau \cap H| }{\pi}
%\le |\tau \cap \F| + |L|\eps \cdot (2^{i-1}p) .
%$$
%Consider first the case $\overline{L}(\tau) \ge 2^{i-1}p$.
Since $L \subseteq \F$, the above inequality can be written as
$$
|\tau \cap \F| - \eps |\tau \cap \F| - \eps \cdot (2^{i-1}p) |\F|
\le \frac{|\tau \cap \F_1| + \pi|\tau \cap H|}{\pi} 
\le |\tau \cap \F| + \eps |\tau \cap \F| + \eps \cdot (2^{i-1}p) |\F|,
$$
%and since $L \subseteq \F$, this becomes
%$$
%|\tau \cap \F|(1 -\eps) \le \frac{|\tau \cap \F_1| + \pi|\tau \cap H|}{\pi} 
% \le |\tau \cap \F|(1 + \eps) ,
%$$
or
\begin{equation}
  \label{eq:my_approx1}
  \frac{|\tau \cap \F|}{|\F|}(1 -\eps) - \eps \cdot (2^{i-1}p) \le 
  \frac{|\tau \cap \F_1| + \pi|\tau \cap H|}{\pi |\F|} 
  \le \frac{|\tau \cap \F|}{|\F|}(1 + \eps) + \eps \cdot (2^{i-1}p) .
\end{equation}
For $i=1, \dots, \log{(1/p)}$, since $\frac{|\tau \cap \F|}{|\F|} \ge (2^{i-1}p)$ (by definition),
this implies that 
$$
\frac{|\tau \cap \F|}{|\F|}(1 - 2\eps) \le 
\frac{|\tau \cap \F_1| + \pi|\tau \cap H|}{\pi |\F|} 
\le \frac{|\tau \cap \F|}{|\F|}(1 + 2\eps) .
$$
When $i=0$, we have $\frac{|\tau \cap \F|}{|\F|} < p$, and then, using similar considerations as above,
%Proposition~\ref{prop:F_1} 
the case $i=0$ of~(\ref{eq:approx_size}) implies:
$$
\frac{|\tau \cap \F|}{|\F|} - \eps p 
\le  \frac{|\tau \cap \F_1| + \pi|\tau \cap H|}{\pi |\F|} 
\le \frac{|\tau \cap \F|}{|\F|} + \eps p ,
$$
and we can replace the term $\eps p$ by $2 \eps p$ in order to be consistent with the previous form
obtained for  $i=1, \dots, \log{(1/p)}$.

Note that the measure approximating $\frac{|\tau \cap \F|}{|\F|}$ can be interpreted to be defined on a \emph{weighted sample},
where each object of $\F_1$ is assigned a unit weight, and each object of $H$ is assigned a (fractional) 
weight $\pi$. Also, observe that the total expected weight of the sample satisfies
$\EE\{|\F_1| + \pi|H| \} = \pi |L| + \pi|H| = \pi|\F|$ (matching the denominator in our measure), and so in our construction
this weighted measure replaces the standard ``uniform measure'' $\overline{Z}(\tau)$ (defined in the introduction) 
%(applied in, e.g.,~\cite{HS-11}) 
resulting when the entire relative approximation is obtained as a uniform sample.

%In the case $\overline{L}(\tau) < 2^{i-1}p$ we proceed as follows.
%Let us assume first that $\frac{|\tau \cap \F|}{|\F|} \ge 2^{i-1}p$, for $i=1, \ldots, \log{(1/p)}$.
%$$
%|\tau \cap \F| - |L|\eps \cdot (2^{i-1}p) 
%\le \frac{|\tau \cap \F_1| + \pi|\tau \cap H|}{\pi} 
%\le |\tau \cap \F| + |L|\eps \cdot (2^{i-1}p) .
%$$
%By our assumption we obtain:
%$$
%|\tau \cap \F| - |L| \eps \cdot \frac{|\tau \cap \F|}{|\F|} 
%\le \frac{|\tau \cap \F_1| + \pi|\tau \cap H|}{\pi} 
%\le |\tau \cap \F| + |L|\eps \cdot \frac{|\tau \cap \F|}{|\F|}  .
%$$
%Using once again the fact that $L \subseteq \F$:
%$$
%|\tau \cap \F| - \eps \cdot |\tau \cap \F|
%\le \frac{|\tau \cap \F_1| + \pi|\tau \cap H|}{\pi} 
%\le |\tau \cap \F| + \eps \cdot |\tau \cap \F| ,
%$$
%which has the same form as~(\ref{eq:my_approx1}).
%
%We are left to address the case $\frac{|\tau \cap \F|}{|\F|} < p$ (referring to layer $\S_0$).
%Since $L \subseteq \F$, we obtain the inequality:
%\begin{equation}
%  \label{eq:my_approx2}
%  \frac{|\tau \cap \F|}{|\F|} - \eps \cdot p \le
%  \frac{|\tau \cap \F|}{|\F|} - \eps \cdot p/2 \le 
%  \frac{|\tau \cap \F_1| + \pi|\tau \cap H|}{\pi |\F|} 
%  \le \frac{|\tau \cap \F|}{|\F|} + \eps \cdot p/2
%  \le \frac{|\tau \cap \F|}{|\F|} + \eps \cdot p .
%\end{equation}

Scaling the parameter $\eps$ appropriately, we conclude:
% that~(\ref{eq:my_approx1}) holds when $\frac{|\tau \cap \F|}{|\F|} \ge p$,
%and~(\ref{eq:my_approx2}) holds otherwise, and thus the (approximate) measure 
%$\frac{|\tau \cap \F_1| + \pi|\tau \cap H|}{\pi |\F|}$ satisfies the desired relative 
%$(p,\eps)$-approximation properties.

\begin{theorem}
  \label{the:approximation}
  Let $(\F, \T)$ be a well-behaved range space.
  Then there exist two subsets $\F_1 \subseteq \F$, $H \subseteq \F$, with the following properties:
  Each range $\tau \in \T$ satisfies:
  $$
  \frac{|\tau \cap \F|}{|\F|}(1 -\eps) 
  \le \frac{|\tau \cap \F_1| + \pi|\tau \cap H|}{\pi |\F|} 
  \le \frac{|\tau \cap \F|}{|\F|}(1 + \eps) ,
  $$
  if $\frac{|\tau \cap \F|}{|\F|} \ge p$, and
  $$
  \frac{|\tau \cap \F|}{|\F|} - \eps p 
  \le \frac{|\tau \cap \F_1| + \pi|\tau \cap H|}{\pi |\F|} 
  \le \frac{|\tau \cap \F|}{|\F|} + \eps p ,
  $$
  otherwise. 
  Thus the sample $\F_1 \cup H$ is a (weighted) relative $(p, \eps)$-approximation for $(\F, \T)$, and its
  overall size is only $O(( \max\{\log\log{(1/p)},\log{\phi(|\F|)} \} + \log{(1/\eps)}) / \eps^2 p)$.
\end{theorem}

Scaling appropriately the parameter $\eps$ once again and also $p$ (see Section~\ref{sec:construction} for the discussion) we obtain:
\begin{corollary}
  \label{cor:small_relative_approx}
  Let $(X, \R)$ be a well-behaved range space.
  Then there exists a (weighted) relative a $(p, \eps)$-approximation for $(X, \R)$, 
  of size $O((\max\{\log\log{(1/p)},\log{\phi(|\F|)} \} + \log{(1/\eps)}) / \eps^2 p)$.
\end{corollary}

\vspace{-2ex}
\paragraph{A polynomial-time algorithm.}
We note that classifying the objects as light or heavy can be done in time polynomial in
$|\F|$ (we omit these straightforward details here).
In order to apply the Local Lemma in a constructive manner, we resort to a recent result by Moser and Tardos~\cite{MT-10}.
In our scenario we apply the extended version of Lemma~\ref{lem:local_lemma} (including the event $\B$),
discussed briefly above and proved in Appendix~\ref{app:proof_sample_size}.
The main property required in order to apply the randomized algorithm described in~\cite{MT-10} is the fact that the 
objects in $\F_1$ are chosen randomly and independently with probability $\pi$. 
That is, the set $L$ induces a finite set of mutually and independent random variables.
Then each event $\A_{\tau} \in \E$, as well as $\B$, is determined by these variables.
Omitting any further details, we obtain:

\begin{theorem}
\label{the:algorithm}
 Given a well-behaved range space $(X,\R)$, one can construct in expected polynomial time, 
 a (weighted) relative $(p, \eps)$-approximation for $(X,\R)$, whose size is
 $O((\max\{\log\log{(1/p)},\log{\phi(|\F|)} \} + \log{(1/\eps)}) / \eps^2 p)$.
\end{theorem}

\vspace{-2ex}
\paragraph{Applications}
%\label{sec:applications}
Using the bound in Theorem~\ref{the:approximation}
%construction in Section~\ref{sec:construction} 
we obtain several geometric settings that admit small-size relative $(p,\eps)$ approximations.
We review these settings and their analysis in Appendix~\ref{app:applications}, and conclude:

%Scaling the parameters $\eps$, $p$ appropriately (see Section~\ref{sec:construction} for the discussion) we obtain:
\begin{corollary}
  \label{cor:boxes_triangles}
  Any range space of points and axis-parallel rectangles in the plane admits a (weighted) relative $(p,\eps)$-approximation 
  of size  $O((\log\log{(1/p)} + \log{(1/\eps)}) / \eps^2 p)$, for any $0 < p < 1$, $0 < \eps < 1$, 
  which can be constructed in expected polynomial time.
  The same asymptotic bound holds for points and axis-parallel boxes in three dimensions, and points
  and $\alpha$-fat triangles in the plane (where $\alpha > 0$ is a constant).
  When $\eps$ is a constant these bounds match the optimal $\Theta(\log\log{(1/p)} / p)$ bound for $p$-nets.
\end{corollary}

%\noindent {\bf Remark:}
%As mentioned in the introduction the previously known bounds on the size of $p$-nets~\cite{AES-10, PT-11} 
%for the three above cases do not necessarily guarantee a relative bounded error.

\begin{corollary}
  \label{cor:halfspaces}
  Any range space of points and halfspaces in two and three dimensions admits a (weighted) relative $(p,\eps)$-approximation 
  of size  $O((\log\log{(1/p)} + \log{(1/\eps)}) / \eps^2 p)$, for any $0 < p < 1$, $0 < \eps < 1$, which can be constructed 
  in expected polynomial time.
\end{corollary}

\begin{corollary}
  \label{cor:small_union_applications}
  Any dual range space defined on 
  (i) pseudo-disks, (ii) $\alpha$-fat triangles, (iii) locally $\gamma$-fat objects
  and points in the plane admits a relative $(p,\eps)$-approximation of size 
  $O((\log\log{(1/p)} + \log{(1/\eps)}) / \eps^2 p)$, for any $0 < p < 1$, $0 < \eps < 1$,
  which can be constructed in expected polynomial time.
\end{corollary}

\vspace{-2ex}
\section{Concluding Remarks.}
\vspace{-2ex}
This study raises several open problems and further improvements, some of which are under on-going research.
First, it is very likely that the sampling scheme that we introduce can be applied over iterations, where
at the $k$th iterations we are given a sample $\F_{k-1}$, from which we extract the sets  $H_{k-1}$, $L_{k-1}$ 
of the corresponding heavy and light objects, and then sample each object from $L_{k-1}$ with probability
$\pi_k := \frac{\max\{\log^{(k+1)}{(1/p)}, \log{\phi(|F_{k-1}|)} \} + \log{(1/\eps)}}{\log^{(k)}{(1/p)} + \log{(1/\eps)}}$,
%(where $\log^{(\cdot)}{\cdot}$ is the iterated $\log(\cdot)$ function),
obtaining $\F_k$;
we stop at iteration $k$ if $\log^{(k+1)}{(1/p)} < \log{\phi(|F_{k-1}|)}$. 
However, this process involves several technical difficulties since (i) the values of $p$, $\eps$ change over iterations,
and (ii) the (weighted) measure becomes somewhat intricate, as it should consist of $\F_k$ and all sets of heavy objects 
collected over all iterations. The author has several initial bounds obtained for this process, 
and she plans to finalize these details in the full version of this paper.
This will tighten the current (probably, suboptimal) bounds stated in Corollaries~\ref{cor:halfspaces} 
and~\ref{cor:small_union_applications}.
% when $\phi(\cdot)$ grows slower than any logarithmic function, and will thus
%yield better bounds in several geometric scenarios, such as dual range spaces of planar regions of nearly-linear union
%complexity and points in the plane (where we believe the bound currently stated in 
%Corollary~\ref{cor:small_union_applications} is suboptimal). 

Another interesting problem that we plan to study is whether the $\log{(1/\eps)}$ factor in the enumerator of our bound
can be removed. Li~\etal~\cite{LLS-01} obtained such an improvement, where they reduced the previously known bound 
$O(\log{(1/(p\eps))}/\eps^2p)$ in~\cite{Haussler-92, Pollard-86} to 
$O(\log{(1/p)}/\eps^2p)$, which they showed to be optimal in the worst case. This improvement is derived by applying
the \emph{chaining} method of Kolmogorov. Roughly speaking, in this technique the standard union bound over a set of
events (defined in some probability space) is replaced by a tighter bound when considering only a relatively small subset
of events, each of which is ``distinct'' in some sense (such a subset is also called an 
``$\eps$-packing''~\cite{Har-Peled-11}).
%We note that this technique is non-trivial, and 
It is a challenging open problem to combine our machinery with the 
chaining method. Specifically, does an $\eps$-packing exist in our scenario? If so, can one apply the Local Lemma
of Lov{\'a}sz on the corresponding events?

Last but not least is the implications of our approach to the bounds on \emph{combinatorial discrepancy}
for well-behaved range spaces. In particular, even just the case of points and halfplane ranges is already challenging.
Har-Peled and Sharir~\cite{HS-11} showed that in such range spaces $(P,\H)$ the discrepancy $\chi(\tau\cap P)$ 
of each range $\tau \in \H$ is only $O(|\tau \cap P|^{1/4} \log{n})$, 
where $n = |P|$. 
%This bound follows by appropriately matching the points in $P$, which follows from a
%construction of a spanning tree (over the points of $P$) with low crossing number.
This property eventually yields the improved bound for relative $(p,\eps)$-approximations in this scenario; recall that 
the improvement in~\cite{HS-11} is with respect to the dependency on $\eps$. If the factor $\log{n}$ in $\chi(\tau\cap P)$
can be reduced to $o(\log{n})$ then this will yield an improvement in the parameter $p$ as well. Nevertheless, we were
unable to apply our technique on this setting so far, due to the differences in our probabilistic model and the one
applied in~\cite{HS-11}. 
%In out model each point is chosen independently by random (which is crucial for the analysis), whereas
%the choice in~\cite{HS-11} is more restrictive and involves dependency among the points.

\vspace{-2ex}
\paragraph*{Acknowledgments.}
The author wishes to thank Micha Sharir for helpful discussions and for his
help in writing this paper.
The author also wishes to thank Boris Aronov and Sariel Har-Peled for several useful comments,
and, in particular, to Sariel's comment regarding the strength of the result for the case of $p$-nets.
The author also thanks Gabor Tardos for a useful discussion about the constructive proof
of the Local Lemma of Lov{\'a}sz.

\appendix
\section{The Construction}
\label{app:construcrtion_proofs}

\vspace{-2ex}
\subsection{Proof of Lemma~\ref{lem:local_lemma}:}
\label{app:proof_local_lemma}
\vspace{-2ex}
%\begin{proof}
  For simplicity of presentation, we bound $\Prob\{\A_{\tau}\}$, for all $\tau \in \S_0$ by
  $\left(\frac{\eps}{\log{(1/p)} \phi(|\F|)}\right)^{B/2}$, which clearly holds by the bound given in the proof of 
  Proposition~\ref{prop:F_1}. Thus~(\ref{eq:chernoff}) holds for all $i=0, \ldots, \log{(1/p)}$.

  Fix a range $\tau$ and a layer $\S_i$.
  We first observe that $0 < \xx(\A_{\tau}) < 1$. The lower bound is trivial. For the upper bound, 
  we obtain from~(\ref{eq:chernoff}) that $\xx(\A_{\tau}) < \left(\frac{e^{4/B} \cdot \eps}{\log{(1/p)} \phi(|\F|) }\right)^{B 2^{i-1}}$,
  which is smaller than $1$ since $B > 1$ (and is chosen to be sufficiently large) and $p \le 1/8$ by our assumption. 

  We next consider all events $\A_{\tau'}$ with $\A_{\tau'} \sim \A_{\tau}$, $\A_{\tau'} \neq \A_{\tau}$.
  Let $i$ be the layer of the range $\tau$, and let $j$ be the layer of $\tau'$.
  The corresponding product $\prod_{\A_{\tau'} \sim \A_{\tau}, A_{\tau'} \neq A_{\tau}} (1 - \xx(A_{\tau'}))$ 
  in the right-hand side of~(\ref{eq:assignment}) can be spelled out for $\tau$ as
  $$
  \prod_{j = 0}^{\log{(1/p)}} \prod_{\A_{\tau'} \sim \A_{\tau}, \A_{\tau'} \neq \A_{\tau} \atop{\A_{\tau'} \in \E_j} } (1 - \xx(A_{\tau'})) .
  $$
  We lower bound separately the sub-products involving layers $\S_j$ with $i \le j$ and layers $\S_j$ with $i > j$. 
  
  %{\bf 1. i = j:}
  %In this case, since $\tau$ is contained in $O(\Delta_i)$ light disks, each of which contains at most $O({\Delta_i}^3)$ 
  %cells (point-ranges), it follows that $\tau$ has a degree-of-dependency at most $O({\Delta_i}^4)$. Thus
  %$$
  %\Pi_{\A_{tau'} \sim \A_{tau}  \atop {\mbox{$\tau$, $\tau'$ lie at layer $i$}} } \ge 
  %\left(1 - \left(\frac{\exp{4/B} \cdot \eps}{\log{(1/p)}}\right)^{B 2^{i-1}}\right)^{{\Delta_i}^4} .
  %$$
  %Using the approximation $(1-y)^{z} \apprx 1 - zy$, where $0 < y < 1$ and $y < z$, and the fact that 
  %$\Delta_i = \frac{C 2^{i-1} \log{(1/(p\eps))}}{\eps^2}$, we obtain that
  %$\Pi_{\A_{tau'} \sim \A_{tau}  \atop {\mbox{$\tau$, $\tau'$ lie at layer $i$}} } \ge 1/D$, where $D > 0$ is a constant that depends 
  %on $B$ (and thus on $C$), and can be made arbitrarily large by chooosing $B$ sufficiently large.
  
  \noindent{\bf (i)  $i \le j$:}
  In this case, since $\tau$ contains at most $O(\Delta_i)$ light objects (as noted above, $\tau$ may also contains at most  
  $O(\Delta_i)$ heavy objects, but they are ignored at that part of the analysis), 
  each of which participates in at most $O(\phi(|\F|){\Delta_j}^{c+2})$ 
  %cells (i.e., cell-ranges) 
  ranges $\tau'$ of $\S_j$, it follows that $\tau$ has a degree-of-dependency at most 
  $\alpha \cdot \phi(|\F|) \Delta_i {\Delta_j}^{c+2}$, for some absolute constant $\alpha > 0$, with the ranges of $\S_j$. 
  We thus obtain:
  $$
  \prod_{\A_{\tau'} \sim \A_{\tau} , \A_{\tau'} \neq \A_{\tau}  \atop {\tau \in \S_i, \tau' \in \S_j, j \ge i} } 
  (1 - \xx(A_{\tau'})) \ge 
  \left(1 - \left(\frac{e^{4/B} \cdot \eps}{\log{(1/p)} \phi(|\F|)}\right)^{B 2^{j-1}}\right)^{\alpha \phi(|\F|) \Delta_i {\Delta_j}^{c+2}} .
  $$
  We simplify the right-hand side using the inequality $(1-y)^{z} \ge 1 - zy$, for $0 < y < 1$, and the fact that 
  $\Delta_k = \frac{C 2^{k-1} \log{(1/(p\eps))}}{\eps^2}$, for each $k = 0, \ldots, \log{(1/p)}$.
  Specifically, we have 
  $$
  \alpha \cdot \phi(|\F|) \Delta_i {\Delta_j}^{c+2} \le \
  \alpha \cdot \phi(|\F|) {\Delta_j}^{c+3} = 
  \alpha' \cdot \phi(|\F|) \frac{2^{(c+3)j} \log^{c+3}{(1/(p\eps))}}{\eps^{2(c+3)}},
  $$
  for another constant $\alpha'$. We then have
  $$
  \left(1 - \left(\frac{e^{4/B} \eps}{\log{(1/p)} \phi(|\F|)} \right)^{B 2^{j-1}}\right)^{\alpha \phi(|\F|) \Delta_i {\Delta_j}^{c+2}}  \ge
  \left(1 - \left(\frac{e^{4/B} \eps}{\log{(1/p)} \phi(|\F|)} \right)^{B 2^{j-1}}\right)^{\alpha \phi(|\F|) {\Delta_j}^{c+3}} 
  $$
  $$
  \ge
  1 - \left(\frac{e^{4/B} \eps}{\log{(1/p)} \phi(|\F|)} \right)^{B 2^{j-1}} \cdot 
  \alpha' \cdot \phi(|\F|) \frac{2^{(c+3)j}\log^{c+3}{(1/(p\eps))}}{\eps^{2(c+3)}} .
  $$
  By ``stealing'' a small portion $\beta \cdot 2^{j-1}$ from the exponent $B \cdot 2^{j-1}$, for some sufficiently
  large constant $\beta > 0$, we can assume that 
  % (recall that $\phi(|\F|)$ does not grow faster than any poly-logarithmic function of $|\F|$):
  $$
  \left(\frac{e^{4/B} \eps}{\log{(1/p)} \phi(|\F|)} \right)^{\beta 2^{j-1}} \cdot 
  \alpha' \cdot \phi(|\F|) \frac{2^{(c+3)j}\log^{c+3}{(1/(p\eps))}}{\eps^{2(c+3)}} < 1 .
  $$
  By choosing $B$ sufficiently large, we thus obtain:
  $$
  \prod_{\A_{\tau'} \sim \A_{\tau}  , \A_{\tau'} \neq \A_{\tau}  \atop {\tau \in \S_i, \tau' \in \S_j, j \ge i} } (1 - \xx(A_{\tau'})) \ge 
  1 - \left(\frac{e^{4/B} \eps}{\log{(1/p)} \phi(|\F|)} \right)^{B' 2^{j-1}} ,
  $$
  for a suitable constant $0 < B' < B$ that 
  %where $B > B' > 0$, is a constant that depends on $B$ (and thus on $C$), and 
  can be made arbitrarily large by choosing $B$ sufficiently large. 
  Having such a choice for $B'$, and recalling that $p \le 1/8$, we obtain that the latter expression is greater than
  $1 - \left(\frac{1}{\kappa} \right)^{2^{j-1}}$, where $\kappa > 0$ is a large constant, 
  whose choice depends on $B'$ (and thus on $B$).
  
  \noindent{\bf (ii) $i > j$:}
  As in the previous case, $\tau$ has a degree-of-dependency at most $\alpha \cdot \phi(|\F|) \Delta_i {\Delta_j}^{c+2}$ 
  with the layers of $\S_j$, and we have:
  $$
  \prod_{\A_{\tau'} \sim \A_{\tau} , \A_{\tau'} \neq \A_{\tau} \atop {\tau \in \S_i, \tau' \in \S_j, j < i} } 
  (1 - \xx(A_{\tau'})) \ge 
  \left(1 - \left(\frac{e^{4/B} \cdot \eps}{\log{(1/p)} \phi(|\F|)}\right)^{B 2^{j-1}}\right)^{\alpha  \cdot \phi(|\F|) \Delta_i {\Delta_j}^{c+2}} .
  $$
  Choosing $B$ sufficiently large, we can assume that $\left(\frac{e^{4/B} \cdot \eps}{\log{(1/p)} \phi(|\F|)}\right)^{B 2^{j-1}} \le 1/2$.
  We can now use the inequality $(1-y)^{z} \ge e^{-2y z}$, for $0 < y \le 1/2$. 
  Proceeding as in the previous case, we have
  $$
  \alpha  \cdot \phi(|\F|) \Delta_i \Delta_j^{c+2} 
  = \alpha' \cdot \phi(|\F|) \frac{2^{i+(c+2)j} \log^{c+3}{(1/(p\eps))}}{\eps^{2(c+3)}} , \quad \mbox{so}
  $$
  $$
  \left(1 - \left(\frac{e^{4/B} \cdot \eps}{\log{(1/p)} \phi(|\F|)}\right)^{B 2^{j-1}}\right)^{\alpha \cdot \phi(|\F|) \Delta_i {\Delta_j}^{c+2}}  
  \ge
  $$
  $$
  \exp\left\{- 2 \left(\frac{e^{4/B} \cdot \eps}{\log{(1/p)} \phi(|\F|)}\right)^{B 2^{j-1}} \cdot 
  \alpha' \cdot \phi(|\F|) \frac{2^{i+(c+2)j} \log^{c+3}{(1/(p\eps))}}{\eps^{2(c+3)}} \right\} .
  $$
  Applying an exponent stealing similar to the one above, we can cancel most of the other part of the expression, 
  and end up with a lower bound of the form:
  $$
  \prod_{\A_{\tau'} \sim \A_{\tau} , \A_{\tau'} \neq \A_{\tau} \atop {\tau \in \S_i, \tau' \in \S_j, j < i} } 
  (1 - \xx(A_{\tau'})) \ge 
  \exp\left\{- 2^{i} \cdot \left(\frac{e^{4/B} \eps}{\log{(1/p)} \phi(|\F|)} \right)^{B'' 2^{j-1}} \right\},
  $$
  where $0 < B'' < B$ is another constant sufficiently close to $B$.
  We can in fact assume that $B'' = B'$ by replacing the larger of them by the smaller.
  Choosing $B$ (and thus $B'$) to be sufficiently large, and using, as above, the inequality $p \le 1/8$, 
  we obtain that
  $$
  \prod_{\A_{\tau'} \sim \A_{\tau} , \A_{\tau'} \neq \A_{\tau}  \atop {\tau \in \S_i, \tau' \in \S_j, j < i} } 
  (1 - \xx(A_{\tau'})) \ge 
  \exp\left\{- 2^{i} \cdot \left(\frac{1}{\kappa} \right)^{2^{j-1}} \right\},
  $$
  where $\kappa > 0$ can be chosen to be the same constant as in case (i). 
  %a large constant, whose choice depends on $B'$ (and thus on $B$).

  \vspace{-2ex}
  \paragraph{Putting the bounds together:}
  We thus obtain, for a fixed range $\tau$,
  $$
  \prod_{j = 0}^{\log{(1/p)}} \prod_{\A_{\tau'} \sim \A_{\tau}, \A_{\tau'} \neq \A_{\tau} \atop{\tau' \in \S_j} } 
  (1 - \xx(A_{\tau'})) \ge
  \prod_{j=0}^{i-1} \exp\left\{- 2^{i} \cdot \left(\frac{1}{\kappa} \right)^{2^{j-1}} \right\} 
  \cdot \prod_{j=i}^{\log{(1/p)}} \left(1 - \left(\frac{1}{\kappa} \right)^{2^{j-1}}\right) .
  $$
  For the first product we have
  $$
  \prod_{j=0}^{i-1} \exp\left\{- 2^{i} \cdot \left(\frac{1}{\kappa} \right)^{2^{j-1}} \right\}  = 
  \exp\left\{- 2^{i} \sum_{j=0}^{i-1} \left(\frac{1}{\kappa}\right)^{2^{j-1}} \right\} 
  \ge \exp\left\{- 2^{i}/(\sqrt{\kappa} - 1)\right\} .
  $$
  For the latter product we have
  $$
  \prod_{j=i}^{\log{(1/p)}} \left(1 - \left(\frac{1}{\kappa} \right)^{2^{j-1}}\right) 
  \ge 1 - \sum_{j=i}^{\log{(1/p)}} \left(\frac{1}{\kappa}\right)^{2^{j-1}} 
  \ge 1 - \frac{1}{\kappa^{2^{i-1}} - 1}
  %\ge 1 - \frac{D}{D-1} \cdot \left(\frac{1}{D}\right)^{2^{i-1}} 
  \ge 1 - \frac{1}{\sqrt{\kappa} - 1}.
  $$
  This implies that
  $$
  \prod_{j = 0}^{\log{(1/p)}} \prod_{\A_{\tau'} \sim \A_{\tau}, \A_{\tau'} \neq \A_{\tau} \atop{\tau' \in \S_j} } 
  (1 - \xx(A_{\tau'})) \ge
  \left(1 - \frac{1}{\sqrt{\kappa} - 1}\right) \cdot \exp\left\{- 2^{i}/(\sqrt{\kappa} - 1)\right\} .
  $$
  
  Given that $\kappa$ is chosen to be sufficiently large we obtain
  \begin{equation}
    \label{eq:final_condition}
  \exp\left\{2^{i+1} \right\} \cdot \Prob\{\A_{\tau}\} 
  \cdot \prod_{j = 0}^{\log{(1/p)}} \prod_{\A_{\tau'} \sim \A_{\tau}, \A_{\tau'} \neq \A_{\tau} \atop{\tau' \in \S_j} } 
  (1 - \xx(A_{\tau'})) 
  %\ge \exp\left\{2^{i-1}/2 \right\} \times \Prob\{\A_{\tau}\} 
  \ge \Prob\{\A_{\tau}\} ,
  \end{equation}
  as asserted.
  This completes the proof of the lemma.
%\end{proof}

\vspace{-2ex}
\subsection{Proof of Corollary~\ref{cor:F_1}}
\label{app:proof_sample_size}
\vspace{-2ex}
To this end, 
%it is sufficient to show that (we omit the straightforward details leading to this form):
%$$
%\Prob\{ |\F_1| > (1+\gamma) \EE \{|F_1|\} \} < \bigwedge_{i=0}^{\log{(1/p)}} \bigwedge_{\tau \in \S_i } (1 - \Prob\{\A_{\tau}\}) ,
%$$
%for some sufficiently large constant $\gamma > 0$.
%This will show that such a sample exists with a positive probability.
%However, we show a slightly stronger property, in which 
%
we strengthen~(\ref{eq:all_events}) to include the event $\B:|\F_1| > (1+\gamma) \EE \{|F_1\}$, 
where $\gamma$ is a sufficiently large constant.\footnote{
  If fact, it is sufficient to show that $\Prob\{ \B \}$ is smaller than
  $\bigwedge_{i=0}^{\log{(1/p)}} \bigwedge_{\tau \in \S_i } (1 - \Prob\{\A_{\tau}\})$. However, in our analysis
  we include the event $\B$ into the local lemma in order to eventually be able to apply it in a constructive manner.
}
%and then strengthen~(\ref{eq:all_events}) to include $\B$ as well.
That is, we show that there exists an additional assignment $\xx' : \B \rightarrow (0,1)$, such that
\begin{equation}
  \label{eq:all_events_B}
  (1 - \Prob\{\B\}) \bigwedge \left\{\bigwedge_{A_{\tau} \in \E} (1 - \Prob\{\A_{\tau}\}) \right\}
  \ge (1 - \xx'(\B)) \prod_{A_{\tau} \in \E} (1 - \xx(\A_{\tau})) > 0 ,
\end{equation}
implying that all complementary events occur with a positive probability.
By construction, for each each event $\A_{\tau} \in \E$, we have $\A_{\tau} \sim \B$.
We thus need to modify~(\ref{eq:assignment}) as follows:
\begin{align}
  \label{eq:assignment_B}
  \Prob\{\A_{\tau}\} \le \xx(\A_{\tau}) \cdot (1 - \xx'(\B)) \prod_{\A_{\tau'} \sim \A_{\tau}, \A_{\tau'} \neq \A_{\tau}} (1 - \xx(A_{\tau'})) , &
  \quad \mbox{for each $\A_{\tau} \in \E$, and}
\end{align}
$$
\Prob\{\B\} \le \xx'(\B) \cdot \prod_{\A_{\tau} \in \E} (1 - \xx(A_{\tau})) .
$$
 
Indeed, we obtain using Chernoff's bound once again, the fact that $\EE\{|\F_1|\} = \pi |L|$, and $|L| \ge |\F|/2$:
\begin{align}
  \label{eq:size_F_1}
  \Prob\{\B \} = \Prob\{ |\F_1| > (1+\gamma) \EE \{|F_1|\} \} & 
  < \exp \left\{- \gamma^2/3 \cdot \EE\{|\F_1|\} \right\}
\end{align}
\begin{align*}
  \le \exp \left\{-\frac{D \gamma^2}{6} \cdot \frac{\max\{\log\log{(1/p)},\log{(\phi(|\F|)} \} + \log{(1/\eps)} }{\eps^2 p}  \right\} .
\end{align*}
%On the other hand, 
By~(\ref{eq:all_events}) and the assignment of $\xx(\A_{\tau})$ in Lemma~\ref{lem:local_lemma} we obtain:
$$
\bigwedge_{i=0}^{\log{(1/p)}} \bigwedge_{\tau \in \S_i } (1 - \Prob\{\A_{\tau}\}) \ge
\prod_{\A_{\tau} \in \E } \left(1 - \xx\{\A_{\tau}\} \right) =
\prod_{\A_{\tau} \in \E } \left(1 - 2^{i+1} \cdot \Prob\{\A_{\tau}\} \right) .
$$
Using~(\ref{eq:chernoff}) and the fact that there are only $O(|\F|\phi(|\F|) \Delta_i^c)$ ranges in layer $i$,
the latter term is lower bounded by 
$$
\prod_{i=0}^{\log{(1/p)}} 
\left(1 - \left(\frac{e^{4/B} \cdot \eps}{\log{(1/p)} \phi(|\F|)}\right)^{B 2^{i-1}} \right)^{\alpha|\F| \phi(|\F|) \Delta_i^c} 
$$
$$
\ge
\prod_{i=0}^{\log{(1/p)}}
\exp\left\{- 2\left(\frac{e^{4/B} \cdot \eps}{\log{(1/p)} \phi(|\F|)}\right)^{B 2^{i-1}}\cdot 
\alpha' \phi(|\F|) \frac{2^{c i}\log^{c+1}{(1/(p\eps))} }{\eps^{2(c+1)} p}  \right\} ,
$$
for two absolute constants $\alpha, \alpha' > 0$.
Applying once again ``exponent stealing'' similarly to the proof of Lemma~\ref{lem:local_lemma}, 
%and using the fact that $\phi(\cdot)$ does not grow faster than any poly-logarithmic function,
we obtain that this term is at least
$$
\exp\left\{ - \sum_{i=0}^{\log{(1/p)}} 
  \left(\frac{e^{4/B} \cdot \eps}{\log{(1/p)} \phi(|\F|)}\right)^{B' 2^{i-1}} \cdot \frac{1}{p} \right\} 
\ge
\exp\left\{ - \left(2\left(\frac{e^{4/B} \cdot \eps}{\log{(1/p)} \phi(|\F|)}\right)^{B'/2} \cdot \frac{1}{p} \right) \right\} 
$$
for a suitable constant $0 < B' < B$ that can be made arbitrarily large by choosing $B$ sufficiently large. 
Comparing the latter exponent with that of~(\ref{eq:size_F_1}), we conclude that
$$
\exp \left\{-\frac{D \gamma^2}{3} \cdot \frac{\log\log{(1/p)} + \log{(1/\eps)} }{\eps^2 p}  \right\} \ll 
\exp\left\{ - \left(2\left(\frac{e^{4/B} \cdot \eps}{\log{(1/p)} \phi(|\F|)}\right)^{B'/2} \cdot \frac{1}{p} \right) \right\} ,
$$
when $\gamma$ is chosen to be sufficiently large. We now put $\xx'(\B)$ to be
$$
\exp \left\{
\left(2\left(\frac{e^{4/B} \cdot \eps}{\log{(1/p)} \phi(|\F|)}\right)^{B'/2} \cdot \frac{1}{p} \right) 
- 
\frac{D \gamma^2}{6} \cdot \frac{\max\{\log\log{(1/p)},\log{(\phi(|\F|)} \} + \log{(1/\eps)} }{\eps^2 p} 
\right\} .
$$
%and then set $\xx'(\B) :=  M \cdot \Prob\{\B\}$.
In other words, we set $\xx'(\B)$ to be the ratio between the upper bound on $\Prob\{\B\}$ and the lower bound
on $\prod_{\A_{\tau} \in \E } \left(1 - \xx\{\A_{\tau}\} \right)$. Due to this property we always have
$\xx'(\B) \cdot \prod_{\A_{\tau} \in \E } \left(1 - \xx\{\A_{\tau}\} \right) > \Prob\{\B\}$, and thus the second part 
of Inequality~(\ref{eq:assignment_B}) is satisfied.
We also note that when $\gamma$ is chosen to be sufficiently large, we should have $\xx'(\B) < 1$.
In fact, $\xx'(\B)$ can be made arbitrarily small (by choosing $\gamma$ sufficiently large), and then we have, say, $\xx'(\B) \le 1/2$.
%
%In order to show the first part of~(\ref{eq:assignment_B}), 
%it is easy to verify that in fact, 
%we observe that since the coefficient $M$ can be made arbitrarily small (by chossing 
%$\gamma$ sufficiently large), we always have, say, $(1 - \xx'(\B)) > 1/2$.
This implies that we can modify Inequality~(\ref{eq:final_condition}) from the
proof of Lemma~\ref{lem:local_lemma} so that it now satisfies 
$$
\exp\left\{2^{i+1} \right\} \cdot \Prob\{\A_{\tau}\} 
\cdot (1 - \xx'(\B)) \prod_{j = 0}^{\log{(1/p)}} \prod_{\A_{\tau'} \sim \A_{\tau}, \A_{\tau'} \neq \A_{\tau} \atop{\tau' \in \S_j} } 
  (1 - \xx(A_{\tau'})) 
  %\ge \exp\left\{2^{i-1}/2 \right\} \times \Prob\{\A_{\tau}\} 
  \ge \Prob\{\A_{\tau}\} ,
$$
which shows the first part of~(\ref{eq:assignment_B}).

\vspace{-2ex}
\subsection{Applications}
\label{app:applications}
\vspace{-2ex}

\subsubsection{Points and Axis-Parallel Boxes in Two and Three Dimensions}
\vspace{-2ex}
We begin with the two-dimensional case.
It is well known that a set $P$ of $n$ points in the plane admits $\Theta(n^2)$
rectangular ``empty'' ranges (that is, these ranges do not contain any point of $P$ in their
interior); see, e.g.,~\cite{AES-10}. 
%In general, the bound on the number of rectangular ranges of size at most $k$ is $O(k^2 n^2)$ (we omit
%the straightforward details followed by the random sampling theory of Clarkson 
%and Shor~\cite{CS-89}). 
Thus we resort to the technique presented in~\cite{AES-10, EHR-12} instead.
We use the following property shown in~\cite{EHR-12} (and based on the 
analysis in~\cite{AES-10}):

\begin{lemma}[Ene~\etal~\cite{EHR-12}]
  \label{lem:rectangles}
  Given a set $P$ of $n$ points in the plane and a parameter $k > 0$, 
  one can compute a set $R_{k}$ of $O(k^2 n \log{n})$ axis-parallel rectangles
  (each of which is ``anchored'' either on its right side or its left side to a vertical line), 
  such that for any axis-parallel rectangle $r$, if $|r \cap P| \le k$, then there exists two rectangles 
  $r_1, r_2 \in R_{k}$ such that $|(r_1 \cup r_2) \cap P| = |r \cap P|$.
\end{lemma}
%
%we consider only ``anchored'' rectangular ranges. That is, 
%rectangles whose left edge or right edge lies on a common vertical line.
%Let us assume, without loss of generality, that these rectangles are anchored
%at their left side.
%Following the analysis in~\cite{AES-10}, the number of such empty
%rectangular ranges is only $O(n)$, and using a standard (balanced) binary tree construction
%over the points in $P$, . In fact, each of these ranges 
%is defined by a triple of points, bounding its respective top, botton, and right edges.
%Following the random sampling theory of Clarkson and Shor, the number $R_{\le k}$ of such ranges
%of size at most $k$ is $O(k^3 \R_0(n/k))$, where $\R_0(m)$ denotes the number of empty
%ranges induced by $m$ points, which is, by the bound just stated, is $O(m\log{m})$.
%We thus obtain that $R_{\le k} = O(k^2 n \log{(n/k)})$ (that is, 
Thus in this case $c=2$ and $\phi(\cdot)$ is the $\log(\cdot)$ function, 
so the range space $(P,R_{n})$ is well-behaved.
In order to bound the relative error for the original (non-anchored) rectangular ranges we proceed as follows.
Let $\tau$ be such a range realized by a rectangle $r$, let $r_1$, $r_2$ be its two corresponding portions 
satisfying the property in Lemma~\ref{lem:rectangles}, and let $\tau_1$, $\tau_2$ be $\{r_1 \cap P\}$, $\{r_2 \cap P\}$,
respectively. We now replace $P$ by the corresponding sample $\F$, and, as before, denote by $\T$ the set of the resulting ranges, 
and then apply Theorem~\ref{the:approximation} in order to obtain:
$$
\frac{|\tau_i \cap \F|}{|\F|}(1 -\eps) - \eps p
\le 
\frac{|\tau_i \cap \F_1| + \pi|\tau_i \cap H|}{\pi |\F|} 
\le \frac{|\tau_i \cap \F|}{|\F|}(1 + \eps) + \eps p,
$$
for $i=1,2$.
By combining these inequalities for $\tau_1$, $\tau_2$, we obtain: 
$$
\frac{|\tau \cap \F|}{|\F|}(1 -\eps) - 2\eps p
\le 
\frac{|\tau \cap \F_1| + \pi|\tau \cap H|}{\pi |\F|} 
\le 
\frac{|\tau \cap \F|}{|\F|}(1 + \eps) + 2\eps p,
$$
This implies that $\F_1 \cup H$ is a relative $(p, 3\eps)$-approximation for $(\F, \T)$ (in the above ``weighted sense''),
and $|\F_1 \cup H| =  O((\log\log{(1/p)} + \log{(1/\eps)}) / \eps^2 p)$.
%Substituting $\eps$ by $\eps/3$ we obtain a relative $(p, \eps)$-approximation of the same asymptotic bound.
%(We omit the easy details).

When $P$ is a set of points in three dimensions, one can obtain similar properties to those in 
Lemma~\ref{lem:rectangles}, derived from the analysis in~\cite{AES-10}.
In this case one can compute a set $B_{k}$ of $O(k^2 n \log^3{n})$ axis-parallel boxes
 % (each of which is ``anchored'' either on its right side or its left side to a vertical line), 
such that for any axis-parallel box $b$, if $|b \cap P| \le k$, then there exists eight boxes 
$b_1, \ldots b_8 \in B_{k}$ such that $|(\cup_{i=1}^8 b_i) \cap P| = |b \cap P|$.
We omit the straightforward details in this version.

Thus in this case $c=2$ and $\phi(\cdot)$ is the $\log^3(\cdot)$ function, so the range space $(P,B_{n})$ is well-behaved.
As above, put $\tau_i := \{b_i \cap P\}$, $i=1, \ldots, 8$.
Replacing $P$ with $\F$ once again, applying Theorem~\ref{the:approximation},
and then combining the resulting inequalities for $\tau_i$, $i=1, \ldots, 8$:
$$
\frac{|\tau \cap \F|}{|\F|}(1 -\eps) - 8\eps p
\le 
\frac{|\tau \cap \F_1| + \pi|\tau \cap H|}{\pi |\F|} 
\le 
\frac{|\tau \cap \F|}{|\F|}(1 + \eps) + 8\eps p,
$$
and thus $\F_1 \cup H$ is a relative $(p, 9\eps)$-approximation for $(\F, \T)$, and its size
has the same asymptotic bound as in the two-dimensional case.

\vspace{-2ex}
\paragraph{Points and $\alpha$-fat triangles in the plane.}
When $P$ is a set of $n$ points in the plane and the ranges are \emph{$\alpha$-fat} triangles 
(that is, triangles for which each of their angles is at least $\alpha$) the analysis
in~\cite{EHR-12} implies:
\begin{lemma}[Ene~\etal~\cite{EHR-12}]
  \label{lem:triangles}
  Given a set $P$ of $n$ points in the plane, a parameter $k$, and a constant $\alpha > 0$,
  one can compute, in polynomial time, a set $\T_{k}$ of $O(k^3 n \log^2{n})$ regions, such that for any
  $\alpha$-fat triangle $\Delta$, if $|\Delta \cap P| \le k$, then there exists (at most) $9$ regions 
  in $\T_{k}$ whose union has the same intersection with $P$ as $\Delta$ does.
\end{lemma}

Using similar arguments and notation as in the case for axis-parallel boxes 
(here we have $c=3$ and $\phi(\cdot)$ is the $\log^2(\cdot)$ function), 
we obtain:
$$
\frac{|\tau \cap \F|}{|\F|}(1 -\eps) - 9\eps p
\le 
\frac{|\tau \cap \F_1| + \pi|\tau \cap H|}{\pi |\F|} 
\le 
\frac{|\tau \cap \F|}{|\F|}(1 + \eps) + 9\eps p,
$$
and thus is $\F_1 \cup H$ is a relative $(p, 10\eps)$-approximation for $(\F, \T)$, and its size
has the same asymptotic bound as in the previous cases.

We now appropriately scale the parameters $\eps$, $p$, for each of the above three settings,
in order to obtain Corollary~\ref{cor:boxes_triangles}.
%Corollary~\ref{cor:boxes_triangles} now follows by scaling the parameters $\eps$, $p$ appropriately 
%(see Section~\ref{sec:construction} for the discussion).

\vspace{-2ex}
\subsubsection{Points and Halfspaces in Two and Three Dimensions}
\vspace{-2ex}
Let $P$ be a set of $n$ points, and let $\H$ denote all halfspace ranges.
Using the random sampling theory of Clarkson and Shor~\cite{CS-89}, for any subset of $m$ points of $P$, 
the number of halfspaces of size at most $k$, for any integer parameter $0 \le k \le m$, is $O(mk + m)$ in two dimensions
and $O(m k^2 + m)$ in three dimensions.
%The random sampling theory of Clarkson and Shor~\cite{CS-89} implies then that 
%in the two-dimensional case, the number of ranges of size at most $k$ is $O(nk)$,
%and in three dimension this bound becomes $O(k^2 n)$.
It thus follows that in both cases $(P,\H)$ is a well-behaved range space,
and we can thus apply Theorem~\ref{the:algorithm} for these cases, thereby 
showing Corollary~\ref{cor:halfspaces}.

\vspace{-2ex}
\subsubsection{Planar Regions of Nearly-Linear Union Complexity and Points}
\vspace{-2ex}
Let $R$ be a set of $n$ (closed) connected planar regions, and let $\U(R) = \bigcup R$ denote the union of $R$.
The combinatorial complexity of $\U(R)$ is the number
of vertices and edges of the arrangement $\A(R)$ that appear along $\bd\U(R)$.
For $r \le n$, let $\uu(r)$ denote the maximum complexity of the union
of any subset of $r$ regions in $R$, measured as above.
We assume that $\uu(r)$ is nearly linear, i.e., $\uu(r) \le r\varphi(r)$, where $\varphi(\cdot)$ is a (sublinear)
slowly growing function. % (which we assume, w.l.o.g., to grow not any faster than a polylogarithmic function).

We now consider the dual range space defined on $R$ and all points in the plane.
In fact, it is fairly standard to show that these ranges correspond to all faces in the 
arrangement $\A(R)$, where each range is the subset of regions covering a fixed face of $\A(R)$.
Using a standard application of the Clarkson-Shor technique~\cite{CS-89}, for any subset of $r$ regions of $R$, 
the number of such faces of size at most $k$, for any integer parameter $k > 0$, is $O(k^2 \uu(r/k))$,
or $O(k r \varphi(r/k))$.
Applying Theorem~\ref{the:algorithm}, we obtain:

\begin{corollary}
  \label{cor:small_union}
  Let $R$ be a set of $n$ planar regions such that the union complexity of any $r$ of them is $\uu(r) =  r \varphi(r)$,
  where $\varphi(\cdot)$ is a (sublinear) slowly growing function. 
  Then any dual range space on $R$ and a set of points in the plane
  admits a (weighted) relative $(p, \eps)$-approximation of size 
  $$
  O\left(\frac{\max\{\log\log{(1/p)} ,\log{\varphi(\log{(1/(\eps p))}/\eps^2 p)}\} + \log{(1/\eps)})}{ \eps^2 p}\right),
  $$
  for any $0 < p < 1$, $0 < \eps < 1$.
  This relative approximation can be constructed in expected polynomial time.
\end{corollary}

We now present several standard families with this property, state their union complexity, 
and then apply Corollary~\ref{cor:small_union} for each of these cases in order to conclude
Corollary~\ref{cor:small_union_applications}.

\paragraph{Pseudo-disks.} 
In a set of pseudo-disks, the boundary of any pair of regions are either disjoint or 
cross twice. In this case $\uu(r) =  O(r)$~\cite{KLPS-86}.

\paragraph{$\alpha$-fat triangles.}
Recall that a triangle is $\alpha$-fat if each of its angles is at least $\alpha$. 
In this case $\uu(n) = O(r \log^*{r})$, where the constant of proportionality
depends on $\alpha$~\cite{AdBES-12, EAS-11}.
When the triangles have roughly the same size, this complexity reduces to $O(r)$~\cite{MPSSW-94}.
When on the sides of these triangles is unbounded (in which case the triangles become
\emph{$\alpha$-fat wedges}), this complexity also reduces to $O(r)$~\cite{ERS-94}. 

\paragraph{Locally $\gamma$-fat objects.}
Given a parameter $0 < \gamma \le 1$, an object $o$ is
\emph{locally $\gamma$-fat} if, for any disk $D$ whose center lies in $o$,
such that $D$ does not fully contain $o$ in its interior,
we have $\mathrm{area}(D \sqcap o) \ge \gamma \cdot \mathrm{area}(D)$, where
$D \sqcap o$ is the connected component of $D \cap o$ that
contains the center of $D$. We also assume that
the boundary of each of the given objects has only $O(1)$ locally
$x$-extreme points, and that
the boundaries of any pair of objects intersect
in at most $s$ points, for some constant $s$.
%
%Given a parameter $0 < \gamma \le 1$, an object $o$ is locally $\gamma$-fat if, for any disk $D$ whose center lies in
%$o$, such that $D$ does not fully contain $o$ in its interior, we have $area(D ??? o) \ge \gamma \cdot area(D)$, where
%$D ??? o$ is the connected component of $D cap o$ that contains the center of $D$. We also assume that
%the boundary of each of the given objects has only $O(1)$ locally $x$-extreme points, and that the
%boundaries of any pair of objects intersect in at most $s$ points, for some constant $s$. 
It is then shown in~\cite{AdBES-12} that $\uu(r) = O(r \cdot 2^{O(\log^*{r})})$, where the constant of 
proportionality (in the linear term) depends on $\gamma$.

%Now Corollary~\ref{cor:small_union_applications} follows from Corollary~\ref{cor:small_union}.

\end{document}